\documentclass[12pt,preprint]{aastex}
\shorttitle{Effective Temperatures of Red Supergiants}
\shortauthors{Levesque et al.}
\begin{document}

\title{The Effective Temperature Scale of 
Galactic Red Supergiants: Cool, But Not As Cool As We Thought}

\author{Emily M. Levesque\altaffilmark{1,2}, and Philip Massey\altaffilmark{2}}

\affil{Lowell Observatory, 1400 W. Mars Hill Road, Flagstaff, AZ 86001}
\email{e\_levesq@space.mit.edu, Phil.Massey@lowell.edu}

\author{K. A. G. Olsen\altaffilmark{3}}
\affil{Cerro Tololo Inter-American Observatory, National Optical Astronomy Observatory, Casilla 603, La Serena, Chile}
\email{kolsen@noao.edu}

\author{Bertrand Plez
and Eric Josselin}
\affil{GRAAL, Universit\'{e} de Montpellier II, 34095 Montpellier Cedex 05, France}
\email{Bertrand.Plez@graal.univ-montp2.fr,Eric.Josselin@graal.univ-montp2.fr}

\author{Andre Maeder and
Georges Meynet}
\affil{Geneva Observatory, 1290 Sauverny, Switzerland}
\email{andre.maeder@obs.unige.ch,georges.meynet@obs.unige.ch}

\altaffiltext{1} {Participant, Research Experiences for Undergraduates program, Summer 2004; current address:
Massachusetts Institute of Technology, 77 Massachusetts Avenue, Cambridge, MA 02139}
\altaffiltext{2} {Visiting Astronomer, Kitt Peak National Observatory, National Optical Astronomy Observatory, which is
operated by the Association of Universities for Research in Astronomy,
Inc., under cooperative agreement with the National Science Foundation.}
\altaffiltext{3}{Visiting Astronomer, Cerro Tololo Inter-American Observatory,
National Optical Astronomy Observatory, which is
operated by the Association of Universities for Research in Astronomy,
Inc., under cooperative agreement with the National Science Foundation.}

\clearpage
\begin{abstract}

We use moderate-resolution optical spectrophometry and the new MARCS
stellar atmosphere models to determine the effective temperatures of 74
Galactic red supergiants (RSGs).  The stars are mostly members of OB
associations or clusters with known distances, allowing a critical
comparison with modern stellar evolutionary tracks. We find we can
achieve excellent matches between the observations and the reddened
model fluxes and molecular transitions, although the atomic lines Ca~I
$\lambda 4226$ and Ca~II H and K are found to be unrealistically
strong in the models.  Our
new effective temperature scale is significantly warmer than those in
the literature, with the differences amounting to 400~K for the
latest-type M supergiants (i.e., M5~I).  We show that the newly derived
temperatures and bolometric corrections give much better agreement with
stellar evolutionary tracks.  This agreement 
provides a completely independent verification of our new temperature scale.
The combination of effective temperature and bolometric
luminosities allows us to calculate stellar radii; the coolest and most
luminous stars (KW Sgr, Case 75, KY Cyg, 
HD 206936=$\mu$ Cep)  have radii of roughly
1500$R_\odot$ (7 AU), in excellent accordance with the largest stellar
radii predicted from current evolutionary theory, although smaller than
that found by others for the binary VV Cep and for the peculiar star VY
CMa.  We find that similar results are obtained for the effective
temperatures and bolometric luminosities using only the de-reddened
$V-K$ colors, providing a powerful demonstration of the
self-consistency of the MARCS models.

\end{abstract}


\section{Introduction}
\label{Sec-intro}

Red supergiants (RSGs) are an important but poorly characterized phase
in the evolution of massive stars.  As discussed recently by Massey
(2003) and Massey \& Olsen (2003), stellar evolution models do not
produce RSGs that are as cool or as luminous as those observed.  Such a
discrepancy is not surprising, given the tremendous challenge RSGs
present to evolutionary calculations.  The RSG opacities are uncertain
because of possible deficiencies in our knowledge of molecular
opacities.  The atmospheres of these stars are highly extended, but
in general the models assume plane-parallel geometry.  In
addition, the velocities of the convective layers are nearly sonic, and
even supersonic in the atmospheric layers, giving rise to shocks
(Freytag et al.\ 2002).  This invalidates the mixing-length
assumptions, making the star's photosphere 
very asymmetric and its radius poorly
defined, as demonstrated by recent 
high angular resolution observations
of Betelgeuse (Young et al.\ 2000).

While considering these challenges, we must, however, recognize that
the ``observed" location of RSGs in the H-R diagram is also highly
uncertain, as it requires a sound knowledge of the effective
temperatures of these stars.  For stars this cool (roughly 3000 to 4000
K), the bolometric corrections (BCs) are quite significant (-4 to -1
mags), and these BCs are a steep function of effective temperature.
This makes an accurate effective temperature scale doubly
necessary, as a 10\% error in $T_{\rm eff}$ would lead to a factor of 2
error in bolometric luminosity computed from $V$,
according to the Kurucz (1992) model
atmospheres as described by Massey \& Olsen (2003).  While
interferometric data has provided a good fundamental calibration of red
{\it giants} (see, for example, Dyck et al.\ 1996), there are not
enough nearby red supergiants to employ this method in determining an
effective temperature scale.  Instead, previous scales have relied upon
using broad-band colors to assign temperatures based on the few red
supergiants with measured diameters (Lee 1970, following Johnson 1964,
1966), or upon ``observed" bolometric corrections (from IR
measurements) combined with the assumption of a blackbody distribution
for the continuum (see Flower 1975, 1977).  
However, $(B-V)_o$ is highly
sensitive to the surface gravity of the star due to the increased
effects of line blanketing with lower surface gravities; the effect is
particularly pronounced at $B$ due to the multitude of weak metal lines
in the region; see discussion in Massey (1998).  
In point of fact, the true continuum at
these temperatures (that produced by continuous
absorption) is probably never seen, as noted by Aller (1960). 
White \& Wing (1978) attempt to get around this problem by a novel
scheme involving an 8-color narrow band filter set, which was fit
by a blackbody curve and iteratively corrected to determine uncontaminated
continua.
However, in all such continuum fits there is always some degeneracy
between changes in the effective temperature and changes in the amount
of reddening to be applied to the models; this is particularly
important for the RSGs, as they can be heavily reddened.  

An alternative approach would be to make use of the incredibly rich TiO
molecular bands which dominate the optical spectra of M-type stars.
Atmospheric models, however, have not always included an
accurate opacity, especially for molecular transitions.  The problem
largely stems from the fact that any molecule found in a stellar
atmosphere---even that of an M-type star---must be considered ``high
temperature", while most laboratory data have been obtained at much
lower temperatures and do not include high-excitation transitions.
The situation has decided improved in the
past years, in large part due to the great efforts by a few groups to
compute ab initio line lists (e.g., Partridge \& Schwenke 1997, Harris et
al.\ 2003), and it now seems quite satisfactory regarding oxygen-rich
mixtures, such as TiO (see reviews by Gustafsson \& Jorgensen 1994 and
Tsuji 1986).

The new generation of MARCS models (Gustafsson et al.\ 1975; 
Plez et al.\ 1992)
now includes a much-improved
treatment of molecular opacity (see Plez 2003, Gustafsson et al. 2003).
Using absolute spectrophotometry (both
continuum fluxes and the strengths of the G-band for K stars and the
TiO bands for M stars), these models can now be used to make a far more
robust determination of $T_{\rm eff}$ and the reddening.
Since band strengths are used to
determine the spectral type of RSGs, these new models could serve as a
definitive connection between spectral type and $T_{\rm eff}$.

Here we present spectrophotometry of 74 Galactic 
K- and M-type supergiants (Sec.~\ref{Sec-obs}), 
and use fits of the MARCS models to determine
effective temperatures (Sec.~\ref{Sec-analysis})\footnote{We are making
both the observed spectra and models available to others
via data files at the Centre de Donnees Astronomiques de Strasbourg (CDS).}.
We begin
the analysis by
reclassifying the stars (Sec.~\ref{Sec-types}), and 
then 
construct a new $T_{\rm eff}$ scale for Galactic RSGs (Sec.~\ref{Sec-teff}).
Most of these stars are members of associations
and clusters with known distances, allowing us to also place the stars
on the H-R diagram for comparison with the latest generation of
stellar evolutionary models, which include the effects of stellar rotation (Sec.~\ref{Sec-evolution}).  In Sec.~\ref{Sec-K},
we compare the physical parameters of these stars derived in the optical
to those found from K-band photometry
in order to test the self-consistency of the
MARCS models.

In future work, we will extend this study to the lower metallicity
Magellanic Clouds, where the distribution of spectral subtypes is
considerably earlier than in the Milky Way (Elias et al.\ 1985; Massey
\& Olsen 2003). The lower abundance of TiO may by itself
lead to a lower $T_{\rm eff}$ for stars of the same spectral subtype,
as suggested by Massey \& Olsen (2003).

\section{Observations}
\label{Sec-obs}

\subsection{The Sample}

Our stars are listed in Table~\ref{tab:stars}.  The sample was selected
in order to cover the full range of spectral subtypes from early K
through the latest M supergiants.  The sample was originally chosen to
contain only stars with probable membership in OB associations and
clusters with known distances (Humphreys 1978, Garmany \& Stencel 1992),
but we supplemented this list with
spectral standards from the list of Morgan \& Keenan (1973) in order to
help refine the spectral classification.  We indicate both cluster
membership and/or use as a spectral standard
in the table.  The photometry comes from a variety of sources, as
indicated; one should keep in mind that most (if not all) of these
stars are variable at the level of several tenths or more in $V$.
We have included the $V-K$ colors as well, where the $K$ comes from
Josselin et al.\ (2000) and references therein.

Distances of OB associations are notoriously uncertain, as most are
far too distant for reliable trigonometric parallaxes; instead, spectral
parallaxes need to be used, often resulting 
in a large dispersion due to the large scatter in $M_V$ for a given spectral
subtype (see Conti 1988).  In Table~\ref{tab:dists} we give the
distances from several sources, when possible, for each OB association or
cluster in our sample; in general, the agreement is within a few tenths
of a magnitude.  We also include in Table~\ref{tab:dists} the ``average"
reddenings determined from the values for {\it early}-type members
listed by Humphreys (1978) for the OB associations. For the clusters, we
list the average reddening values given by Mermilliod \&
Paunzen (2003).  We note that in general membership in an
OB association is never perfectly well established, and therefore there
is an additional uncertainty connected with the distance of any particular
star.

\subsection{Spectrophotometry}

The spectrophotometry data were obtained during three observing runs: two with 
the KPNO 2.1-meter telescope and GoldCam Spectrograph
(17-24 March 2004 and 28 May - 1 June 2004), 
and one with the CTIO 1.5-meter telescope and Cassegrain CCD spectrometer 
(7-12 March 2004). Similar resolutions and wavelength coverage were
obtained in both hemispheres. Detailed information on the observing parameters
is given in Table~\ref{tab:obs}.

The Kitt Peak observations were taken under
sporadically cloudy conditions in March, while all of
the nights in May/June were photometric.  Before observing
each object, the spectrograph was manually rotated so that the
slit was aligned with the parallactic angle. 
 We aimed for a S/N of 50 per spectral resolution
element (4 pixels) at the bluest end of the spectrum, with care being taken
not to saturate the detector at the reddest end.
For the brightest stars ($V<6$), we employed
a 5.0~mag or 7.5~mag neutral density filter.
Observations of the flat-field lamps were obtained both with and
without these filters in order to correct for the wavelength
dependence of the transmission of these ``neutral" filters.
Throughout each night we also observed a set of spectrophotometric standards.
The seeing for these observations was 1.2'' to 3'',
with a typical value of 2''.  Exposures of both dome flats and projector
flats were obtained at the beginning of each night; for the red nights,
we also obtained projector flats throughout the night to monitor any shifting
of the fringe pattern that affects the CCD at longer wavelengths.
Wavelength calibration exposures of a He-Ne-Ar lamp were obtained at the
beginning and end of each night.

At CTIO, it was not practical to always observe at the parallactic
angle; instead, each star was observed with a single exposure through a
very wide slit (for good relative fluxes), followed by a series of
shorter exposures obtained through a narrow slit (for good
resolution).  Projector flats were obtained for all wavelength regions
through both the wide and narrow slits. The exposures were typically
obtained in conjunction with a wavelength calibration exposure,
although in practice we found there was little flexure.  Observations
of spectrophotometric standards were obtained throughout the night.

We reduced the data using IRAF\footnote{IRAF is distributed by the
National Optical Astronomy Observatories, which are operated by the
Association of Universities for Research in Astronomy, Inc., under
cooperative agreement with the National Science Foundation.} packages
CCDRED, KPNOSLIT and CTIOSLIT.  We used dome flats to flatten the blue
Kitt Peak data, and the projector lamps to flatten the red Kitt Peak
data.  We found that the only shift in the fringe pattern in the red
occurred during the thirty minutes following a refill of the dewar with
liquid nitrogen. After that the fringe pattern was stable, so we simply
combined these projector flats.  The spectra were all extracted using
an optimal-extraction algorithm.  When reducing the CTIO data the
narrow slit width observations were combined and divided into the wide
slit observations. The resulting division was fit with a low-order
function and used to correct the fluxes of the narrow-slit
observations.  In a few cases the wide-slit observations included the
presence of an additional star in the extraction aperture; these
stars were eliminated from further consideration, and do not appear in
Table~\ref{tab:stars}.

The observations of spectrophotometric standards were 
used to create sensitivity functions; typically a grey-shift was applied
to each night's data, and the resulting scatter was 0.02~mag.
In addition, the different wavelength regions were grey-shifted to
agree in the regions of overlap.

When we began our analysis we found significant discrepancies between the
reddened model atmospheres and the observed spectra in the near-UV region
($<$4000\AA\ for the reddest stars.  Careful investigations suggest that,
despite the excellent agreement of the spectrophotometric standards,
there could be contamination by a grating ghost in the near-UV, in which
a small fraction of the red light contaminates the data.  The expected
flux ratio $F_{\lambda 7000} / F_{\lambda 3500}$ is roughly 10,000 for the
most reddened M supergiants, 
a regime seldom encountered in astronomical spectrophotometry.  We do not
know if this contamination is a small fraction of the near-UV light, or 
if it dominates, and we have therefore restricted our study to the data
longwards 4100\AA, where our tests show that 
the ghosting effect is negligible.

\section{Analysis}
\label{Sec-analysis}

\subsection{Reclassification of Spectral Type}
\label{Sec-types}

We reclassified each of our stars by visually comparing each spectrum
to the spectral standards.  In order to avoid questions of normalizing
these rich and complicated spectra (with uncertain continuum levels), we
did this comparison in terms of log flux. Given that many of these stars can be
variable in spectral type, we were unsurprised to find that we needed
to reclassify a few of the spectral standards for consistency.  For the
late-K and all M type stars, the classification was based upon the
depths of the TiO bands, which are increasingly strong with later
spectral type.  The reclassification of the early and mid K supergiants
was based primarily on the strengths of the G-band plus the strength
of Ca~I $\lambda 4226$.  These features all get weaker with later
spectral types (Jaschek \& Jaschek 1987), a temperature effect which we
confirmed with the MARCS models.  We list our revised spectral types in
Table~\ref{tab:results}.

\subsection{Modeling the Stars:  Effective Temperatures and Reddenings\label{Sec-teff}}

We compared the observed spectral energy distribution (SED) of each
star to a series of MARCS stellar atmosphere models. The models used in
our comparisons ranged from 3000~K to 4300~K in increments of 100~K, with
$\log g$ from +1.0 to -1.0 in increments of 0.5~dex.  The choice of
surface gravity was derived iteratively; we began by adopting the $\log
g=-0.5$ models for all of the fits, as this surface gravity was
generally what was expected with the old effective temperature scale
and the resulting placement in the HRD. However, our new
temperatures were a bit warmer than that predicted by the old scale, 
so we re-evaluated all of the fits using $\log g=0.0$,
as this was more consistent with the revised locations in the HRD.
The effective temperatures remained unchanged with the use of these
higher surface gravities (except for the occasional
star), but the values of $E(B-V)$ increased slightly.  Finally, we used
the revised temperature scale and bolometric luminosities with the
evolutionary tracks (Sec.~\ref{Sec-evolution}) to compute $\log g$
star-by-star for the RSGs with distances.  This confirmed that our
choice of the $\log g=0.0$ models was appropriate for most of the
stars.  We refit the stars using the appropriate $\log g$ if the
distance was known; otherwise, we adopted $\log g=0.0$.  In practice
the choice of $\log g$ affected the derived values of $E(B-V)$ by
$<0.1$~mag, and had no effect on the derived effective temperatures.

\begin{figure}
\includegraphics[scale=.35,angle=90]{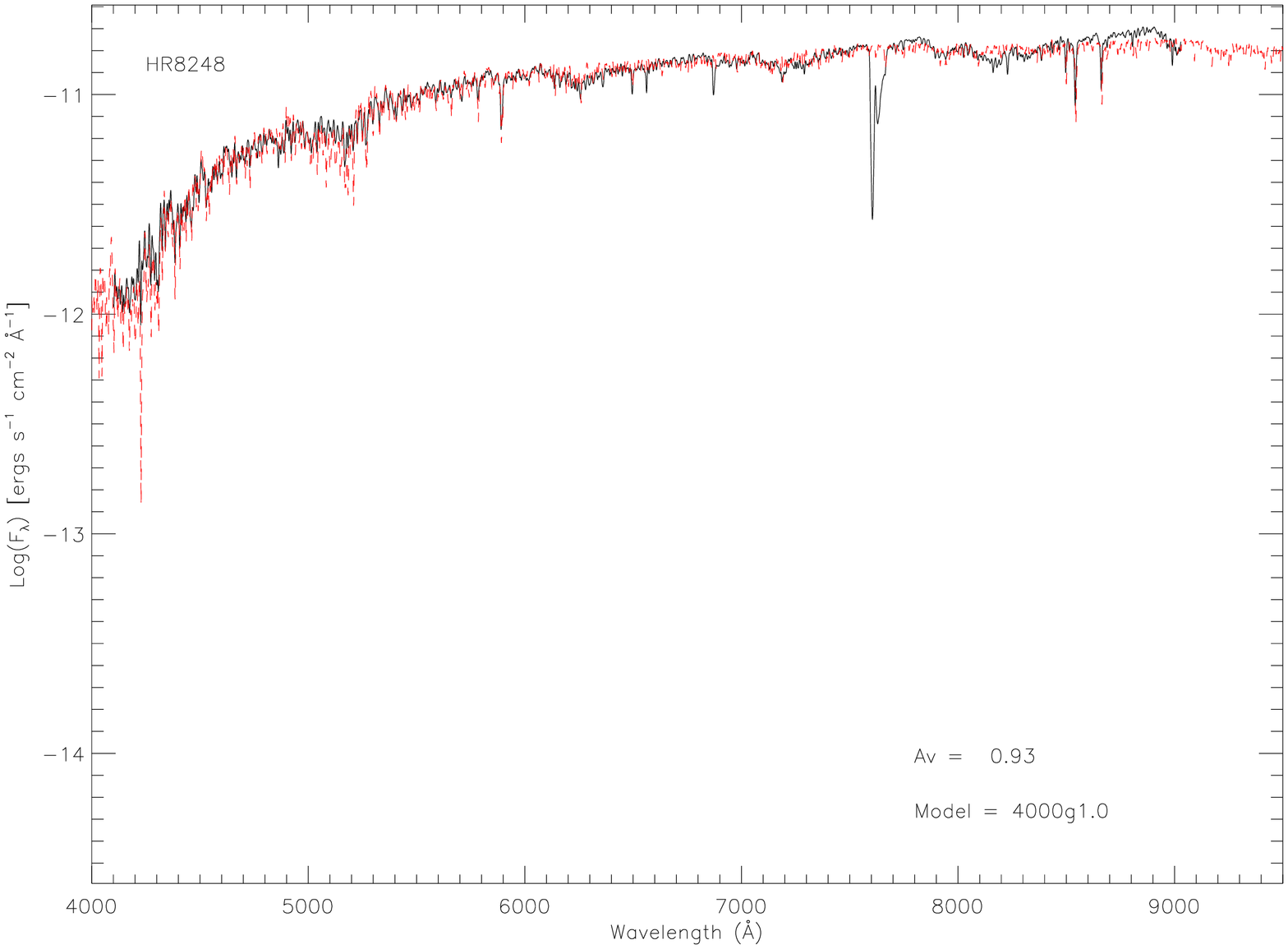}
\includegraphics[scale=.35,angle=90]{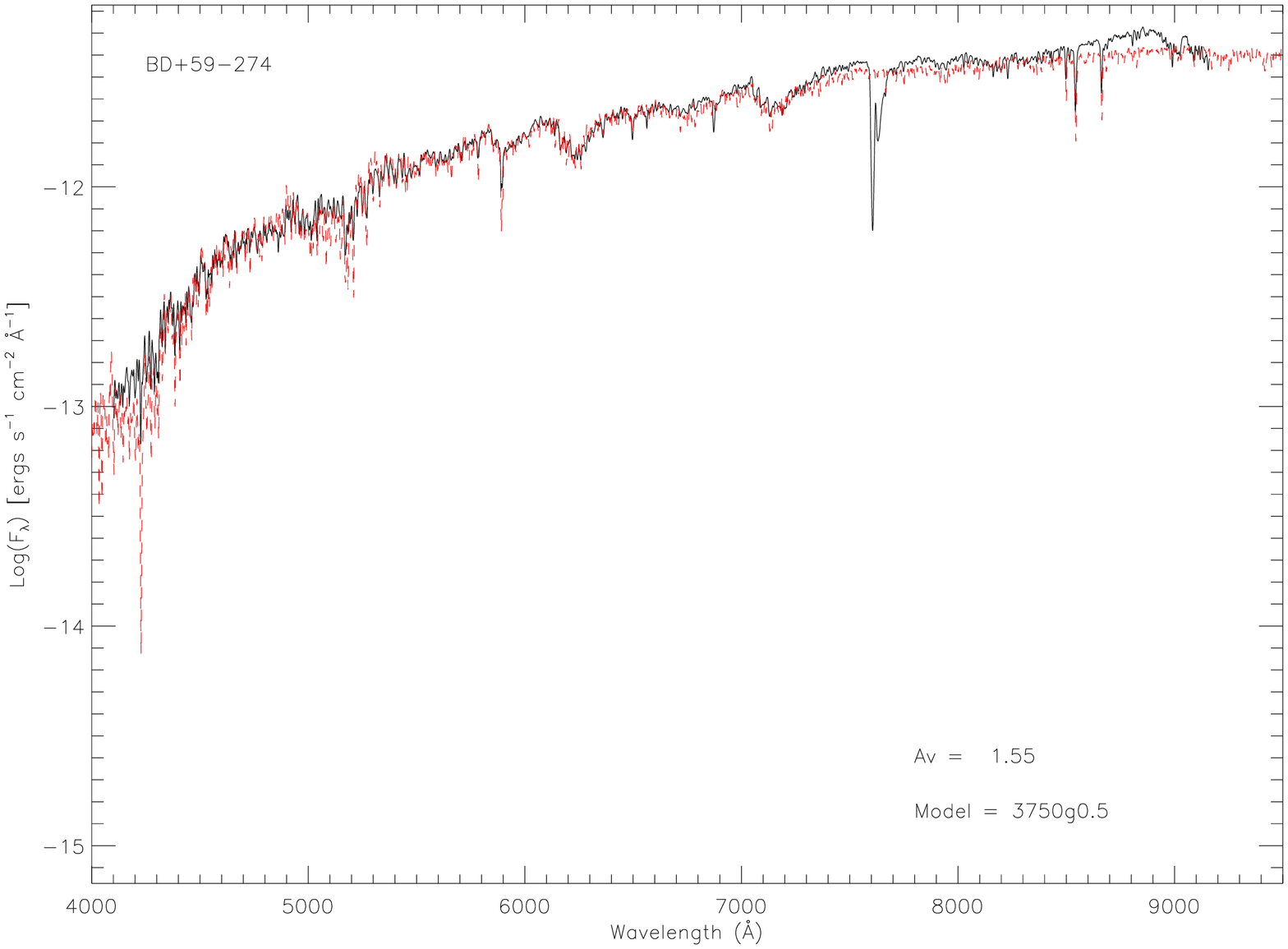}
\includegraphics[scale=.35,angle=90]{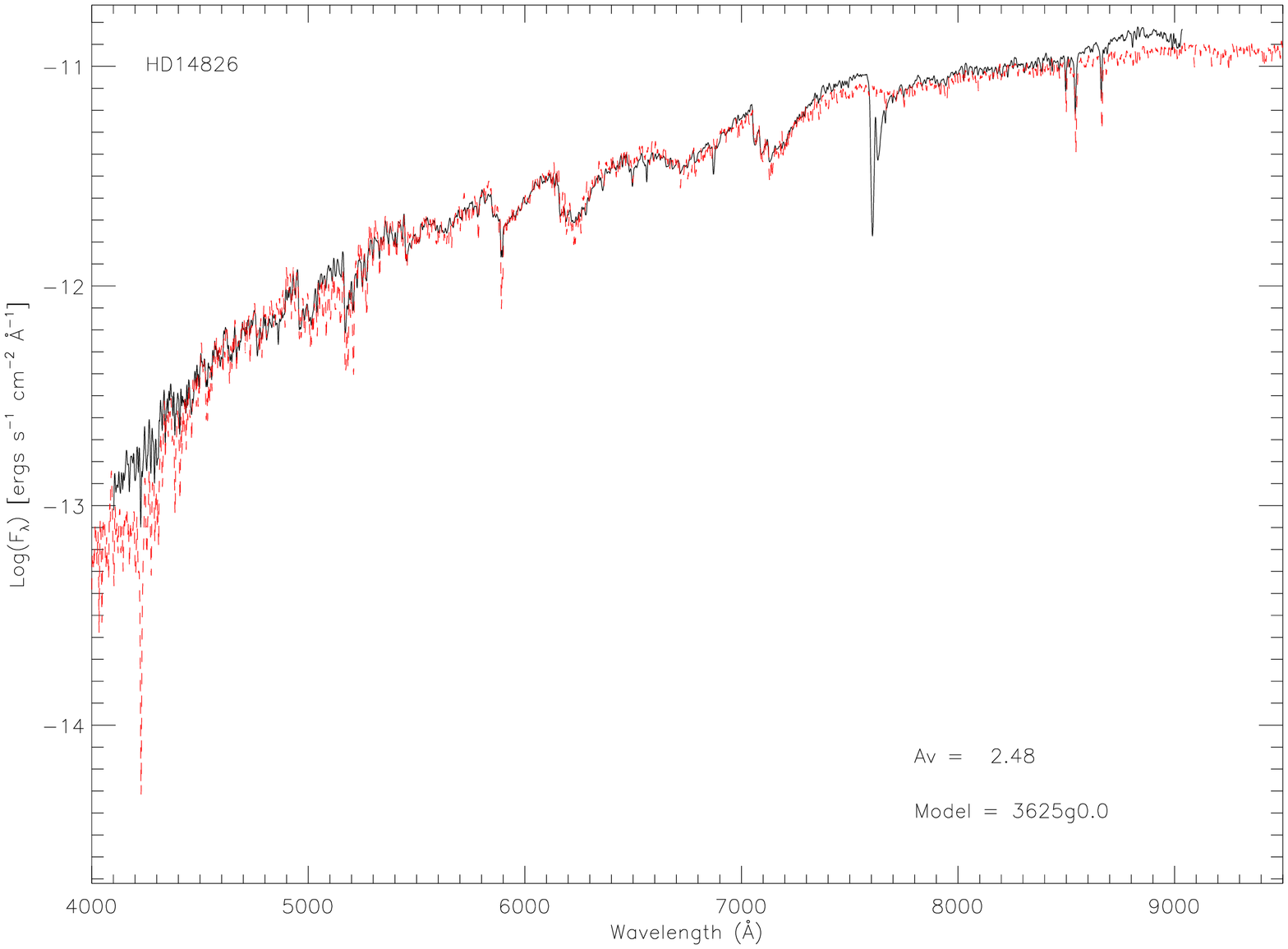}
\includegraphics[scale=.35,angle=90]{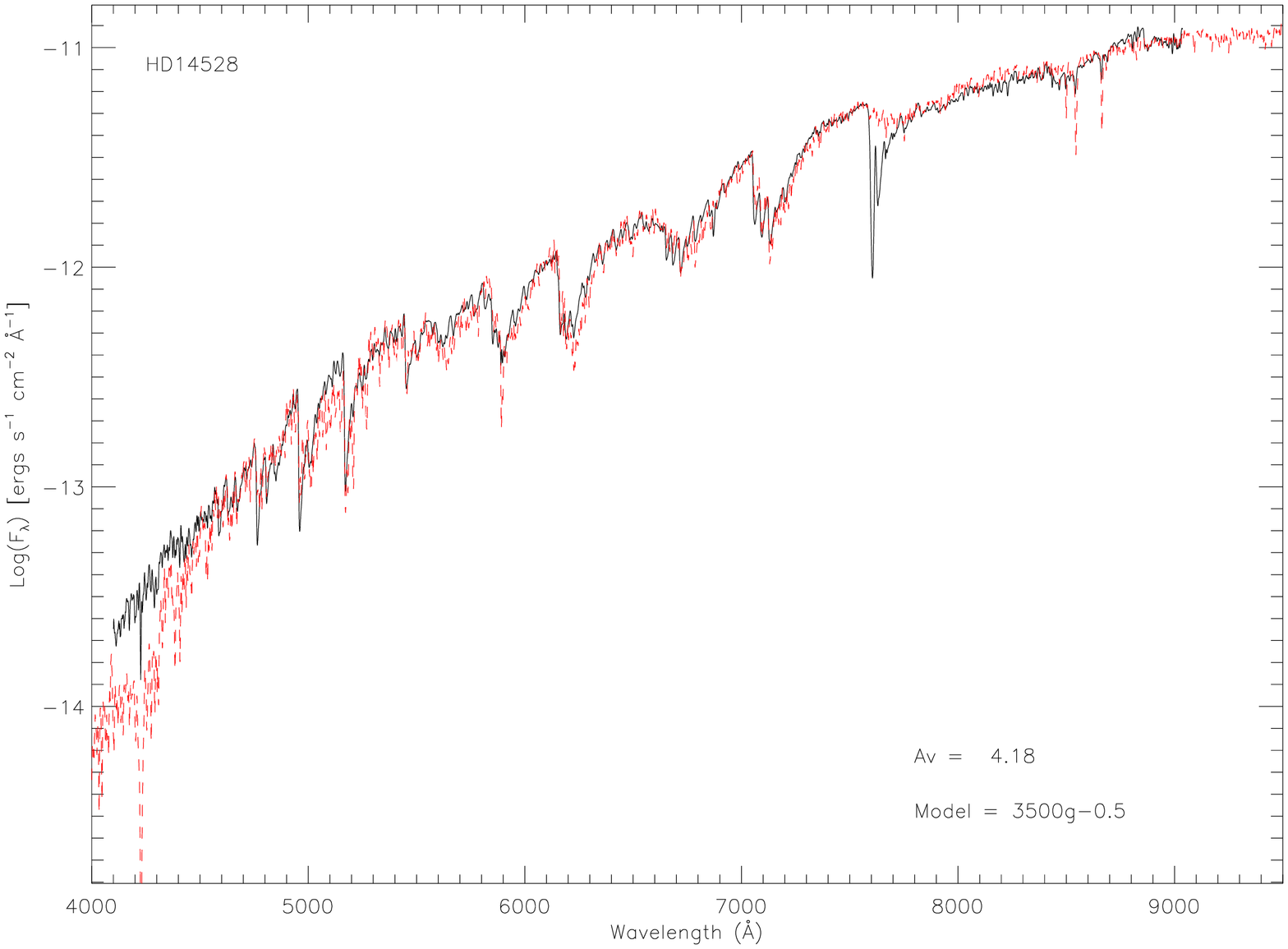}
\caption{\label{fig:plots} A comparison between our spectrophotometry (solid
black)
and the MARCS models (dotted red).  The data are plotted on a $\log F_\lambda$
scale to facilitate comparison between the size of the molecular transitions.
The models have been reddened by the indicated amount using the standard
$R_V=3.1$ reddening law of Cardelli et al.\ (1989).  The stars shown
have spectral types K1~I (HR~8248), M1~I (BD+59 274), M2~I (HD 14826),
and M4.5~I (HD 14528).  The complete set of comparisons is available
electronically.
}
\end{figure}

In making the fits, we reddened each model using a Cardelli et
al.\ (1989) reddening law with the standard ratio of total-to-selective
extinction $R_V=3.1$\footnote{$R_V$ is typically taken to be
$\sim 3.6$ for RSGs (Lee 1970), due to the increase of 
the effective wavelength of broad-band 
filters with redder stars; see McCall (2004) for a recent discussion.
Our own calculations from the models suggest that $R_V\sim 4.4$
would be more appropriate for $BV$ photometry of RSGs.
Note, however, that in the absence of a peculiar reddening law,
$R_V=3.1$ is an appropriate choice for our study, since our
analysis is not based upon broad-band filter photometry, but rather
upon moderate-resolution (5\AA) SEDs; i.e.,
neither the $B$ nor $V$ filters were 
involved in our extinction determinations.
In order to prevent confusion, we use
the equivalent A$_V$ values rather than $E(B-V)$, as our A$_V$ values
will be directly comparable to those of others, while our $E(B-V)$ values
would be larger than those determined from broad-band
photometry for RSGs.}.
Our initial guess for $E(B-V)$ was based upon the average value for the
cluster, i.e., $E(B-V)_{\rm cluster}$ from Table~\ref{tab:dists}.  The
temperature was determined primarily from the strengths of the TiO
bands (M supergiants) and the G-band (early-to-mid K supergiants), with
$E(B-V)$ then adjusted to produce the best fit to the continuum.  In
Fig.~\ref{fig:plots} we show a sample of spectra and fits, covering a
range of spectral types; the complete set is available in the
electronic edition of the journal.  The fitting was all done in log
units of flux in order to facilitate comparisons of line intensities
without the uncertain process of normalization, as mentioned
previously.

As described in the previous section, our initial modeling revealed a
significant discrepancy between the models and the data in the near-UV
region (3500-4000\AA) for the reddest stars, which we first attributed to
a peculiar reddening law, caused, we argued, by circumstellar dust.
However, since we cannot exclude the possibility that the near-UV data are
affected by an instrumental problem, we will revisit this issue at a later
time when we have better data in the near-UV.  

Otherwise, the agreement
between the SEDs of the reddened models and the data is extremely
good, both in terms of the continua and molecular band depths. The only
significant problem we encountered was for the early-to-mid K stars,
where we expected to use both the Ca~I
$\lambda 4226$ line as well as the G-band in modeling the stars, as
these lines are among the primary classification lines for the early
K's.  However, we found that these Ca~I line was stronger
in the MARCS models' synthetic spectra than in our stars, usually by a
factor of 3 or so.  The Ca~I
line show the same qualitative behavior with effective temperature as
expected (see above discussion), but the absolute line strengths in the
models were too strong.
Since Ca has a very low ionization potential (6.11 eV), most of the Ca
is in the form of Ca II: $\sim$99\% in the model at 4000~K and 98\% in
the model at 3600~K (at an optical depth of 0.01 for the continuum).
Thus, a small over-ionization will strongly impact the line strength,
due to either the effects of non-LTE or a slight error in the model's
temperature structure. The occurrence of cool or hot spots on the
stellar surface could also lead to regions where the ionization
equilibrium is strongly affected. 

In addition to the problem with the Ca~I line, the models also
produce Ca~II H and K lines that are significantly stronger than
those observed, even in those stars for which the fluxes of the
dereddened stars and the stars were in good agreement.  This is not
hard to understand, as
RSGs are expected to have chromospheric
activity, and this is not accounted for in the models.  The presence of
a chromosphere would lead to emission in the H and K lines, resulting
in weaker observed lines than would be the case if the lines were
purely photospheric in origin.  This explanation could be further
investigated by means of high-dispersion spectroscopy.

We relied instead upon the G-band for the fitting of the early and
mid-Ks. We substantiated that we obtained similar temperatures using
the G-band and the TiO bands for the late Ks.  The resulting
temperature fits for the early and mid Ks are therefore less certain,
probably $\pm 100$ K.

We list the effective
temperatures in Table~\ref{tab:results}.  In
general, we found that the data could be matched very well by the
models, {\it both} in line strength and in continuum shape,
and we expect that the effective temperatures
of the M supergiants have been obtained to a precision of 50~K.
The A$_V$ values are determined to a precision of about 0.15~mag.
We note with interest that the derived A$_V$ values of about one-third
our sample are significantly 
higher than the average found from the
OB stars in the same associations and clusters
using the data from Humphreys (1978); indeed, the same conclusion
could have been drawn from the older data given
in that paper.  A possible
interpretation is that this extra extinction is due to circumstellar
dust. We include the $\Delta$A$_V$ values in Table~\ref{tab:results}.
We will revisit this issue in a subsequent paper, once our analysis
of the Magellanic Cloud data (where the foreground reddening is low,
and relatively uniform) is complete.

\begin{figure}
\epsscale{0.6}
\includegraphics[scale=.65,angle=90]{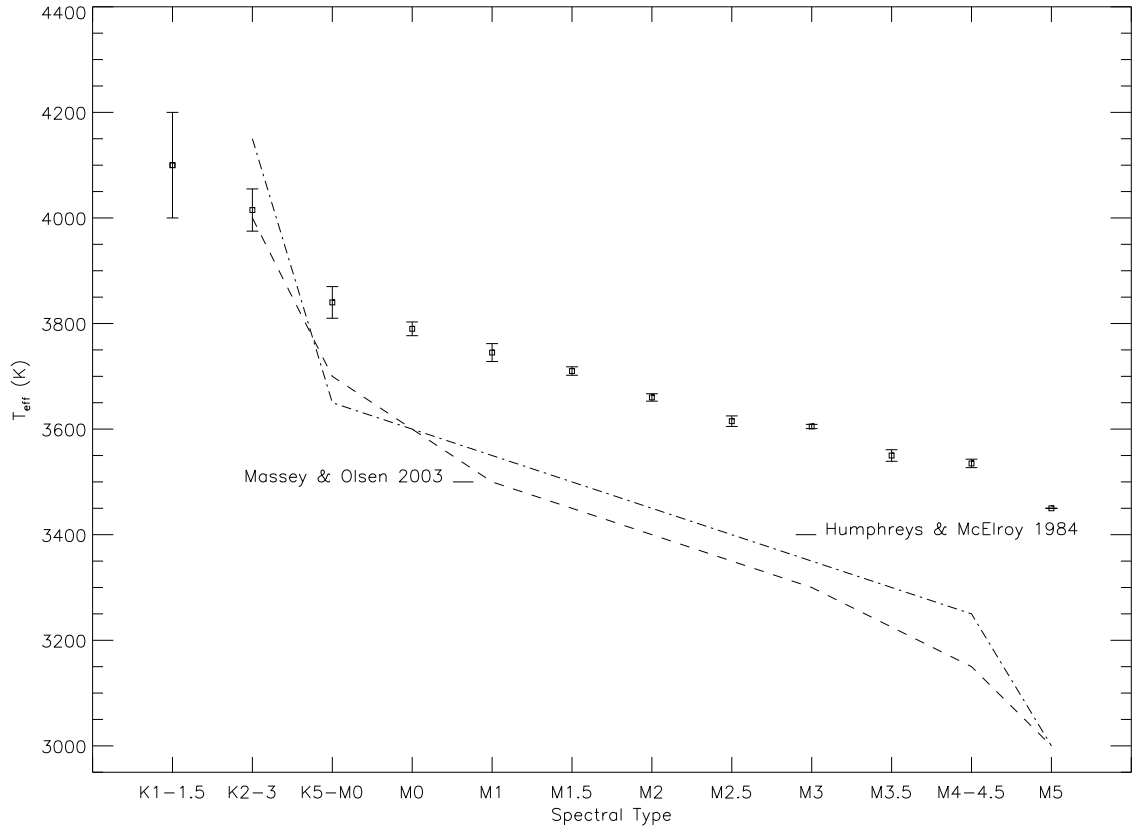}
\caption{\label{fig:tscale} The effective temperature scale for Galactic
RSGs.  The error bars reflect the standard deviation of the means
from Table~\ref{tab:NewT}.  For comparison, we show the
scales of Humphreys \& McElroy (1984) and Massey \& Olsen (2003).
}
\end{figure}

Our new effective temperature scale is given in Table~\ref{tab:NewT},
where we include the number of stars and standard deviation of the mean
($\sigma_\mu$) at each spectral type.  We compare this scale to that of
Humphreys \& McElroy (1984) and Massey \& Olsen (2003) in
Fig.~\ref{fig:tscale}.  Both of the latter are ``averages" from the
literature.  We see that the new scale agrees well for the K supergiants,
but is progressively warmer than past scales for later spectral types,
with the differences amounting to 400~K by the latest M supergiants
(M5~I). The overall progression of the temperature with later spectral
type is more gradual than in past studies.

We include in Table~\ref{tab:NewT} the bolometric corrections to the
V-band corresponding to the new effective temperature scale; these
values are the linear interpolations of the BCs from the MARCS models
given below.  The values are for $\log g=0.0$, but there is little
change with surface gravity ($<0.05$~mag over 0.5~dex in $\log g$ for
$3500\le T_{\rm eff}\le 4300$).
Note that we have referenced the BCs to the system advocated by
Bessell et al.\ (1998), i.e., that the Sun has a BC of -0.07~mag.
This results in values less luminous (by 0.12~mag) than the historical
one; on this system the Sun is {\it defined} to have an $M_{\rm bol}$
of 4.74.

\subsection{Comparison to Evolutionary Models}
\label{Sec-evolution}

In order to compare these stars to the evolutionary models, we must
convert the absolute visual luminosities to bolometric luminosities.
In Table~\ref{tab:BCs} we give the bolometric corrections (BCs) as a
function of effective temperature determined from the MARCS models.  We
list the bolometric magnitudes for each star with a known distance in
Table~\ref{tab:results}\footnote{We note that Josselin et al.\ (2003)
have proposed that the star HD~37536 is an AGB based upon the detection
of Tc and Li.  On the other hand, the star is seen projected against
Aur~OB1, and if membership is assumed, a sensible $M_V$ is derived.  In
addition, the reddening is similar to that of the early-type members of
Aur~OB1.}.

\begin{figure}
\epsscale{0.4}
\plotone{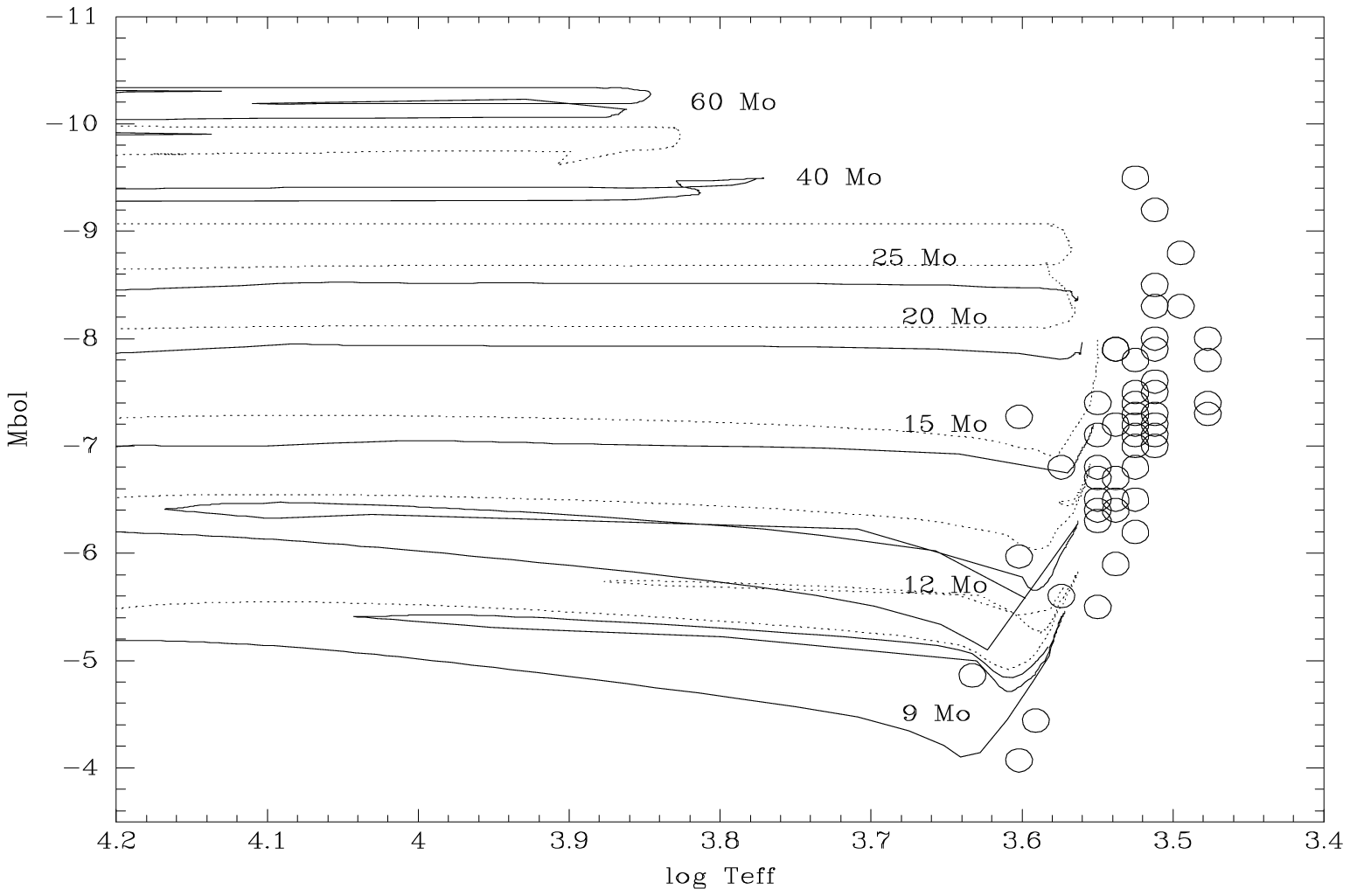}
\plotone{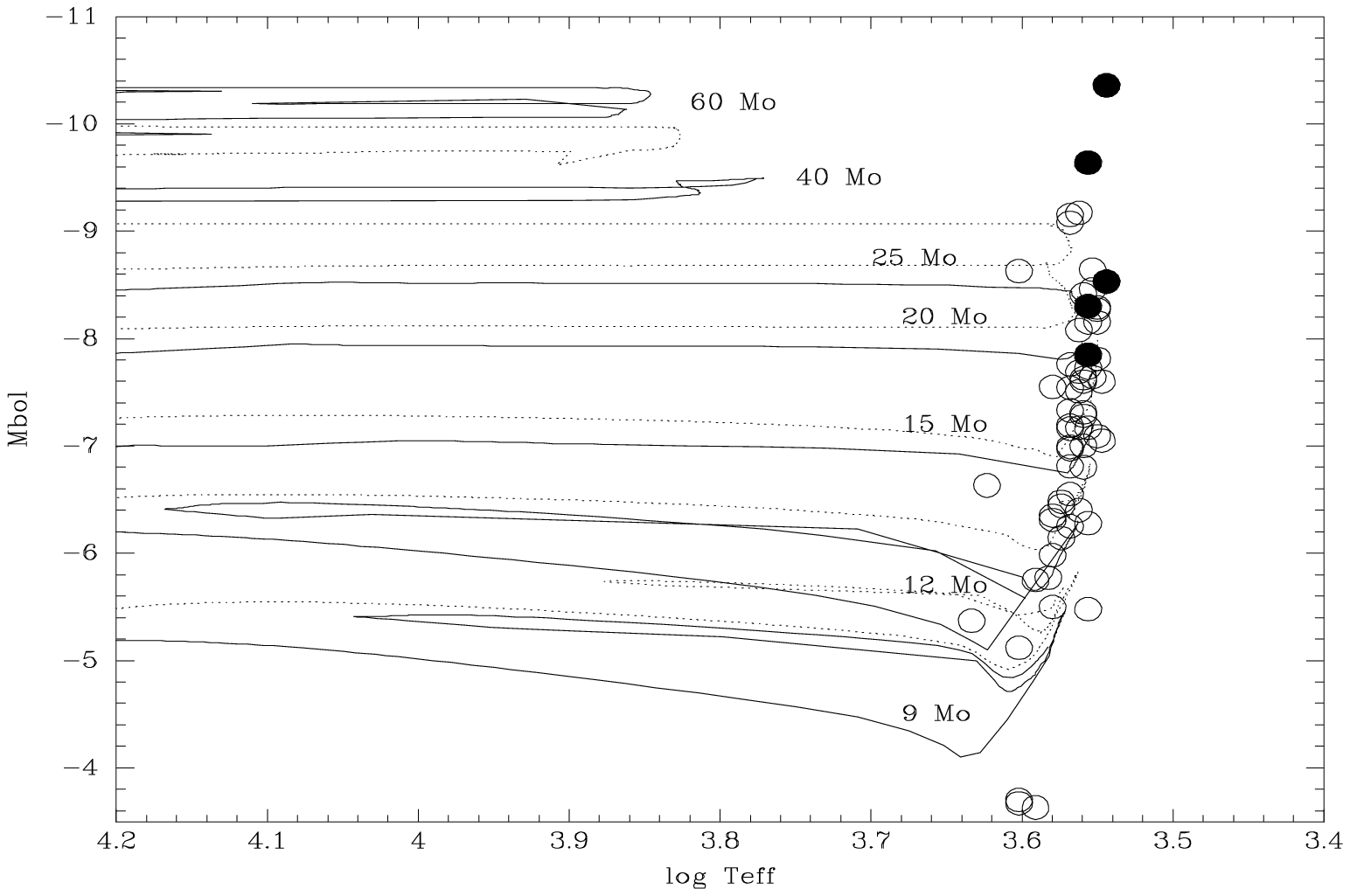}
\plotone{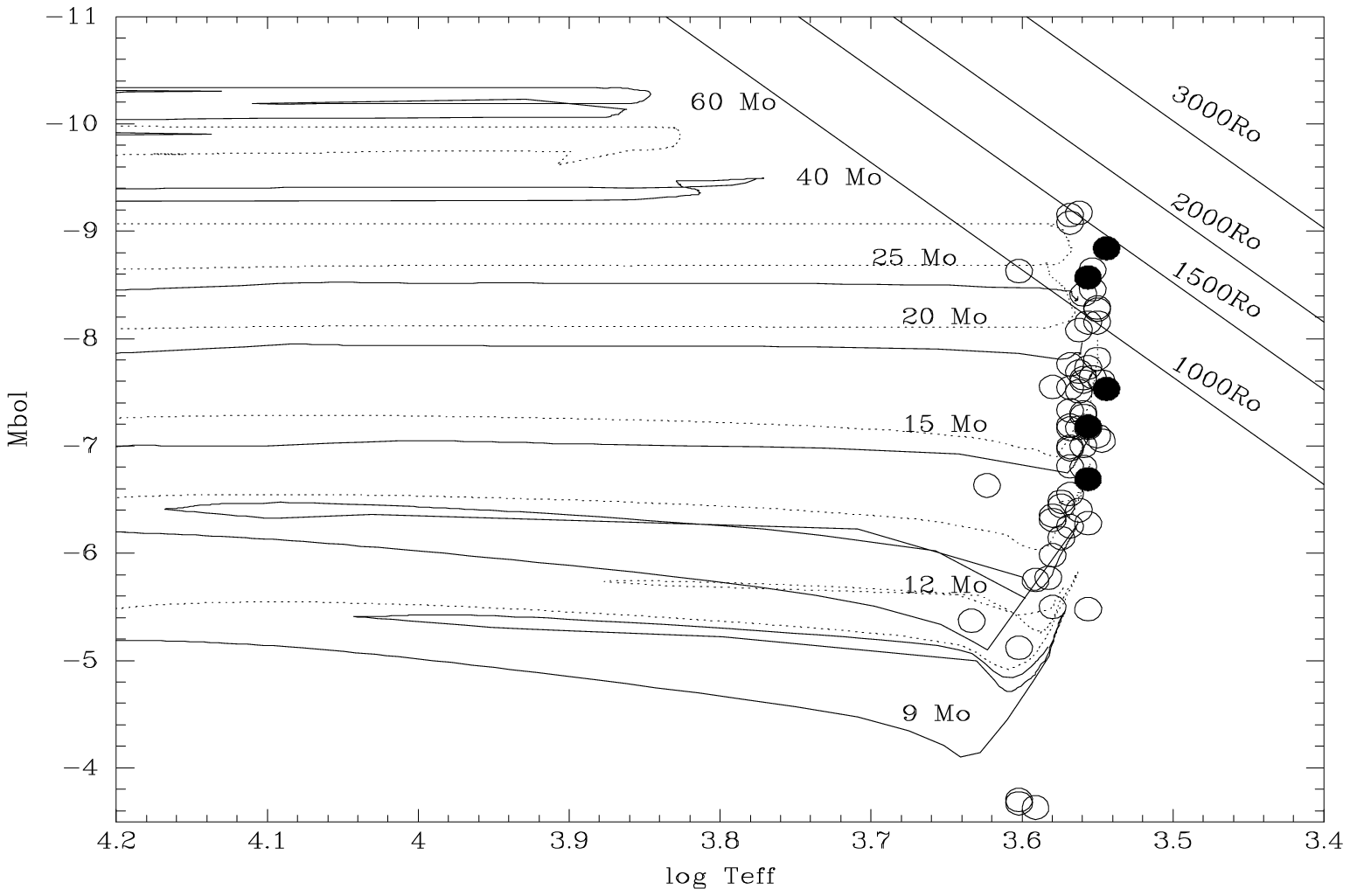}
\caption{\label{fig:HRD} Comparison with evolutionary tracks.
The evolutionary tracks of Meynet \& Maeder (2003) are shown, along
with the location of Galactic RSGs.  The solid lines denote the
no-rotation models, while the dotted lines show the evolutionary
tracks for an initial rotation velocity of 300 km s$^{-1}$.
The tracks with rotation appear above the no-rotation tracks; the
one for 60 $M_\odot$ does not extend this far to the right in the HRD.
In (a) we show the location of the RSGs taken from Humphreys (1978),
using the effective temperature and bolometric corrections of 
Humphreys \& McElroy (1984).  In (b) we show the location of the
RSGs using our new model fits (from Table~\ref{tab:results}). 
The filled circles in (b) denote those five stars whose luminosities
derived from $V$ are significantly higher than those derived from $K$.
In (c) we show these same five stars with the luminosities derived
from $K$.  The diagonal lines at upper right in this figure show lines
of constant radii.
}
\end{figure}

In Fig.~\ref{fig:HRD}(a) we show the solar-metallicity evolutionary
tracks of Meynet \& Maeder (2003) compared to the location of the
Galactic RSGs taken from Humphreys (1978) using the effective
temperature and bolometric corrections of Humphreys \& McElroy (1984).
The disagreement is not as bad as that shown by Massey (2003), as the
new tracks extend further to the right at higher luminosities than did
the older tracks of Schaller et al.\ (1992).  Nevertheless, it is clear
that there are significant differences between theory and
``observation".  In Fig.~\ref{fig:HRD}(b) we now compare the same
tracks to the Galactic RSGs using the new effective temperatures and
bolometric models found by using the MARCS models. 
We have marked with filled symbols those five stars for which the $K$-band
data suggests that our $M_{\rm bol}$ values are too luminous
(Sec.~\ref{Sec-K}), and in Fig.~\ref{fig:HRD}c show the location of these
stars based upon the K-band data.  In (b) and (c) 
the disagreement between theory and location in the HRD has now
disappeared, giving us some confidence in the accuracy of our new
calibration.  A few stars have luminosities significantly higher than
the evolutionary tracks would predict (Fig.~\ref{fig:HRD}b), 
but these are invariably the stars
whose $M_{\rm bol}$ values derived from $V$ are at odds with those derived
from $K$ (Fig.~\ref{fig:HRD}c), presumably due to mistakes
in the correction for reddening, and hence their value must be considered
poorly determined; Fig.~\ref{fig:HRD}c is the best determined.

The fact that we have both M$_{\rm bol}$ and $T_{\rm eff}$ allows us to
determine the stellar radii $R/R_\odot$ from the formal definition of
$T_{\rm eff}$; i.e., $(R/R_\odot)=(L/L_\odot)^{0.5}(T_{\rm
eff}/5781)^2$, where the numerical quantity is the effective
temperature of the sun (see discussion in Bessell et al.\ 1998), and
$L/L_\odot=10^{-(M_{\rm bol}-4.74)/2.5}$.  We include these values in
Table~\ref{tab:results}.  The stars with the
largest radii in our sample, KW~Sgr, Case~75, KY~Cyg, and HD~206936 
(=$\mu$ Cep)\footnote{Also known as Herschel's ``Garnet Star", this
star is often cited as the largest known normal star; see
http://www.astro.uiuc.edu/$\sim$kaler/sow/garnet.html.} 
all
have
radii of roughly 1500$R_\odot$ (7 AU), making these the largest
normal stars known. For comparison, Betelgeuse has a radius
(measured by interferometry) of 645 $R_\odot$ (Perrin et
al.\ 2004).
We include in Fig.~\ref{fig:HRD}(c) lines of constant radii.  These
four large stars are right at what current evolutionary theory predicts
is the maximum radius for Galactic RSGs, as the largest radius
reached by the tracks is found at roughly $M_{\rm bol}=-9$ and 
$\log T_{\rm eff}=3.57$ (3715 K).

Two peculiar RSGs are known with significantly larger radii.
The M2~I primary in the interacting binary VV~Cep has a radius that
has been variously estimated as 1200$R_\odot$ (Hutchings \& Wright 1971) to
1600$R_\odot$ (Wright 1977) and beyond,
with a reasonable upper limit of 1900$R_\odot$ determined
by Saito et al.\ (1980); see
discussion in 
Bauer \& Bennett (2000).  VV~Cep
consists of a RSG primary and a hotter companion orbiting within a
common envelope.  These estimates of the radii are complicated by
the uncertainties in the orbital inclination
in Section 3 of Saito et al.\ 1980), with
the ``definition" of radius determined
by the eclipse method leading to a further ambiguity.
In any event, gravitational interactions are certainly taking place in
this system (Hutchings \& Wright 1971),
and thus may not have applications to normal, single stars.

The other star with a humongous radius is VY CMa.
Using interferometry with Keck, Monnier et al.\ (2004) find
a photospheric radius of 14 AU ($3020 R_\odot$) for this star,
where the distance of 1.5~kpc appears fairly certain from maser 
proper motions (Richards et al.\ 1998; see discussion in Monnier et al.\ 1999).
The properties of this intriguing object have been recently discussed
by Monnier et al.\ (1999) and Smith et al.\ (2001).  With a luminosity
of 2 to 5 $\times 10^{5} L_\odot$ ($M_{\rm bol}=-8.5$ to $-9.5$) 
well established from
the IR (Monnier et al. 1999, Smith 2001), 
the star's temperature would have
to be extremely cool, (2225 K!) to have such a large radius. Using K-band
photometry and a simple model, Monnier et al.\ (2004) suggest an
effective temperature of 2600 K, similar to the 2800 K value 
found by Le Sidaner \& Le Bertre (1996); again, see discussion in
Monnier et al.\ (1999, 2004).
We note that none of these stellar
properties are in accord with stellar evolutionary theory 
(Fig.~\ref{fig:HRD}c),
and, indeed, based upon its inferred mass-loss history,
Humphreys et al.\ (2005) describe the star as ``perplexing",
and argue that it may be in a ``unique evolutionary state."  
Such an object may provide important insight into a previously unrecognized 
avenue of normal stellar evolution, or its peculiarities may be the product
of (for instance) binary evolution, as is the case for VV Cep. Smith et
al.\ (2001) state that any hot, massive companion in VY~CMa would 
have been previously detected spectroscopically, but
we believe that further searching, particularly in the UV, is warranted.
This star was not included in our sample, but we hope to perform an
analysis in the near future.

\subsection{Comparison with $(V-K)_o$}
\label{Sec-K}

In Sec.~\ref{Sec-teff}
we derived a precise relationship between spectral subtype and
effective temperature based primarily on the strengths of the TiO band
strengths.  One test of this result's accuracy is to check for
consistency with other temperature indicators.  Josselin et al.\ (2000)
have emphasized the usefulness of K-band photometry in deriving
bolometric luminosities of RSGs, as the bolometric correction to the
K-band is relatively insensitive to effective temperature and surface
gravities, while RSGs themselves are less variable in the K band than
at $V$.  In addition, correction for interstellar reddening is minor at
$K$.  At the same time, the effective temperature is a very sensitive
function of $(V-K)_o$.  Since over half of our sample have $K$-band photometry
(Table~\ref{tab:stars}), we can perform some
exacting tests of the models to see if we obtained similar physical
parameters by very different techniques.

We have derived synthetic $(V-K)_o$ colors for our models following
the procedure and assumptions of Bessell et al.\ (1998).  The $V$
bandpass comes from Bessell (1990), while the $K$ bandpass comes from
Bessell \& Brett (1988).  Note that the latter is similar to the ``standard"
K system of 
Elias et al.\ (1982) 
and Johnson (1965), and a good approximation to the convolution of detector
and filter used at UKIRT and with most ESO and NOAO instrumentation; 
however, this 
differs considerably from
the $K_s$ filter employed in other
modern instruments, such as the 2Mass survey and the VLT ISAAC
instruments, and the transformations given below will not apply to $K_s$.

\begin{figure}
\epsscale{0.6}
\plotone{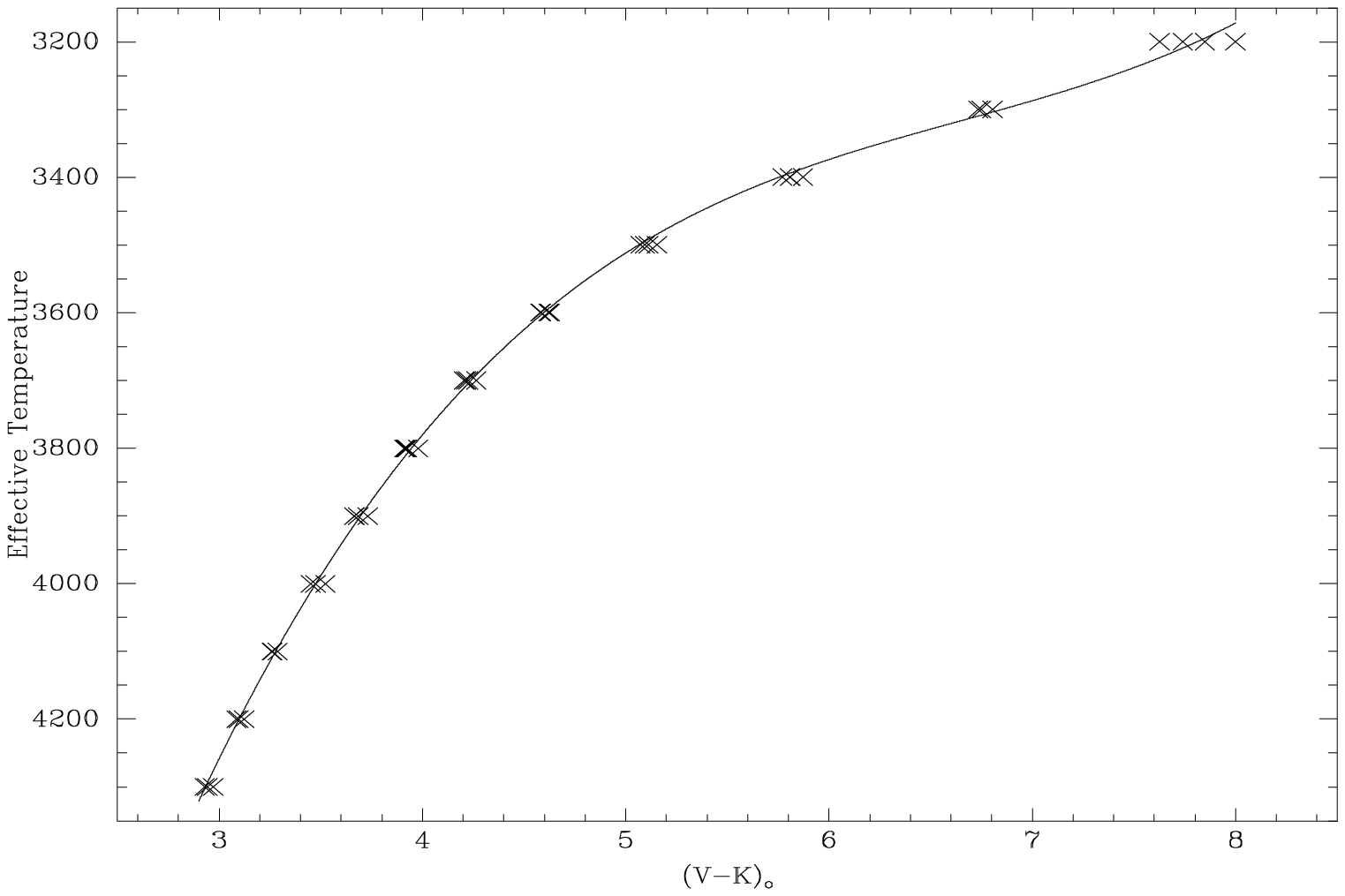}
\plotone{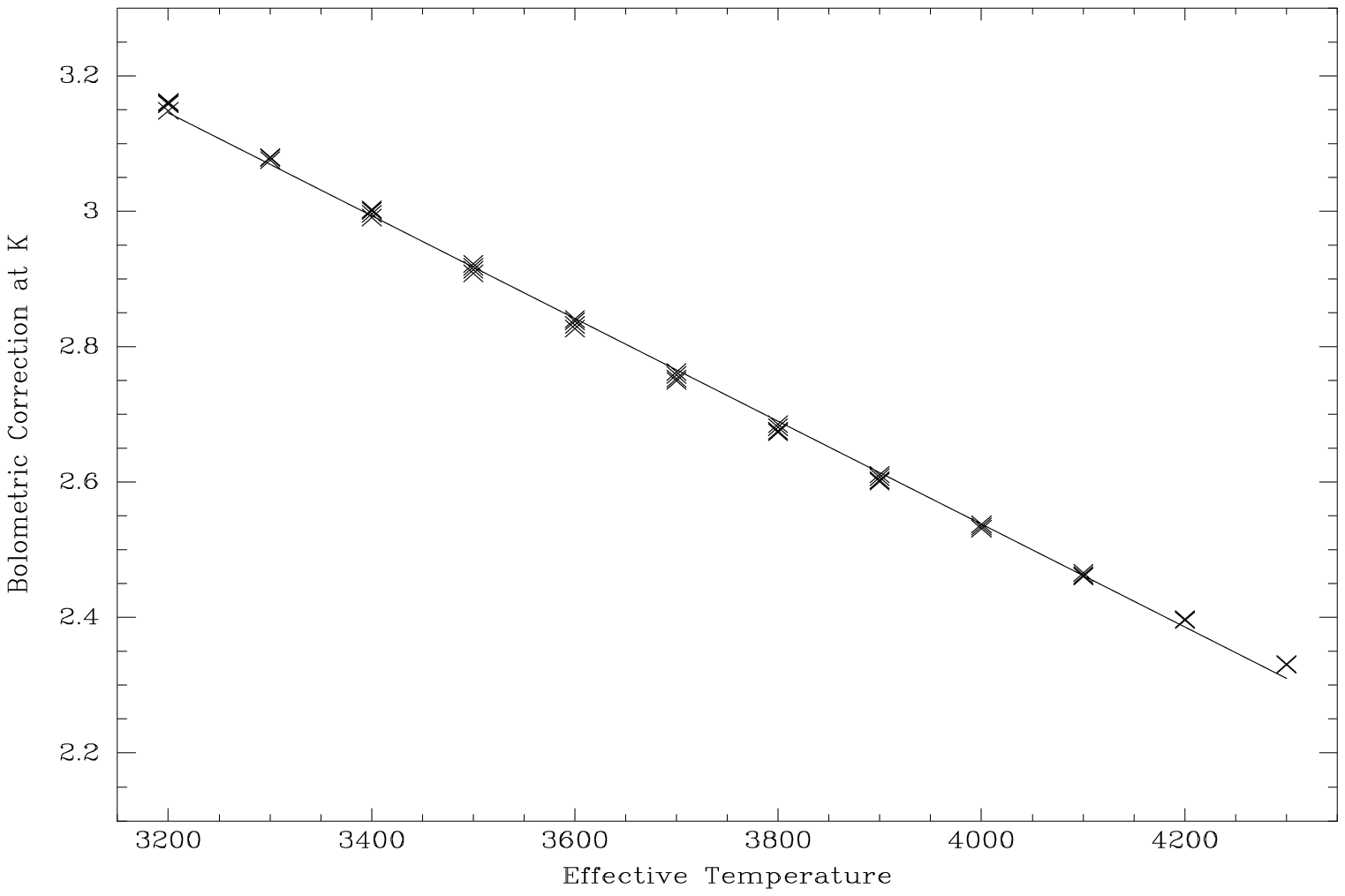}
\caption{\label{fig:K1} Derivation of (a) effective temperatures and (b) bolometric corrections at $K$ based on the $(V-K)_o$ colors. The points were
computed from the models, while the curve shows the smooth fits described
in the text.  The data extends from 3200~K to 4300~K with $\log g$ values
ranging from -1 to +1.
}
\end{figure}

We found that we could approximate the relationship between
$(V-K)_o$ color as a simple power-law: 
$$T_{\rm eff}=7741.9-1831.83 (V-K)_o + 263.135 (V-K)_o^2 -13.1943
(V-K)_o^3,$$
over the range $2.9 <
(V-K)_o < 8.0$, (3200-4300~K) 
with a dispersion of 11K, where the dispersion comes
from considering the full range of appropriate surface gravities
($\log g = -1$ to 1); see Fig.~\ref{fig:K1}a.  The bolometric
correction at K is an almost linear relationship with effective
temperature over the range $3200<T_{\rm eff}<4300$:
$${\rm BC}_K=5.574-0.7589(T_{\rm eff}/1000),$$ with a dispersion of 0.01~mag
(Fig.~\ref{fig:K1}).  We have included these BC$_K$ values in 
Table~\ref{tab:BCs}.

\begin{figure}
\epsscale{0.35}
%\plotone{K2a.eps}
\plotone{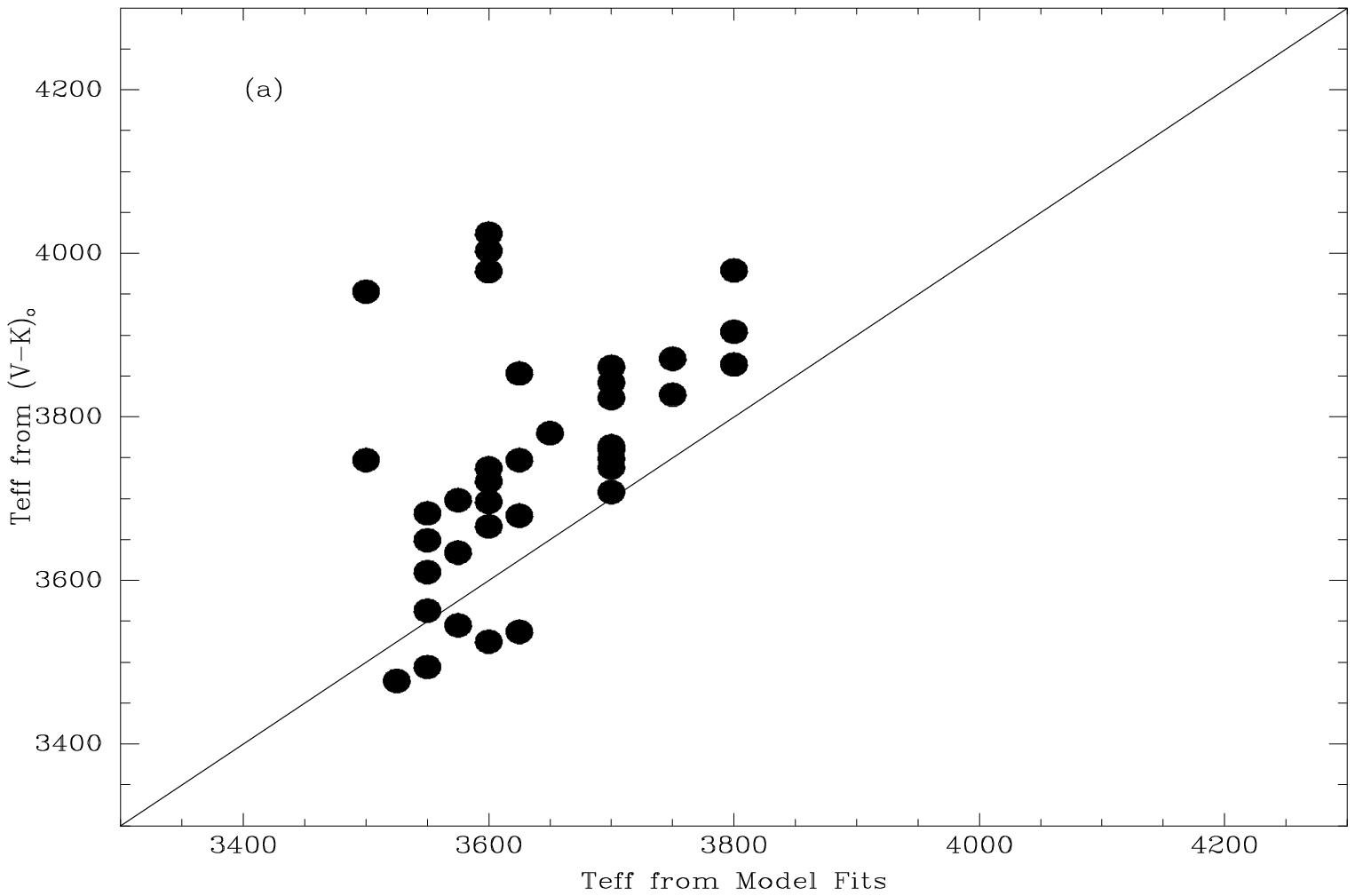}
%\plotone{K2b.eps}
\plotone{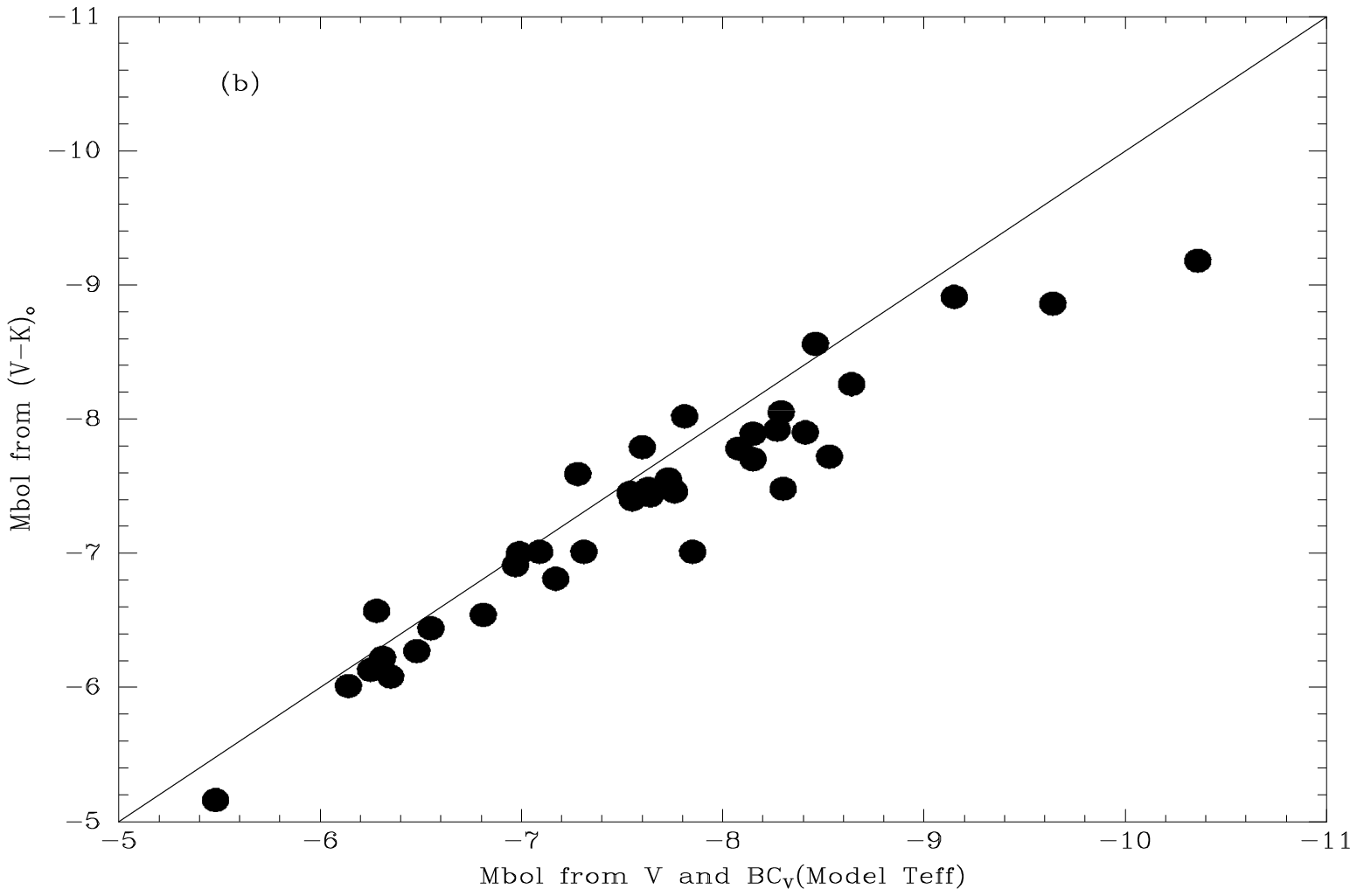}
\plotone{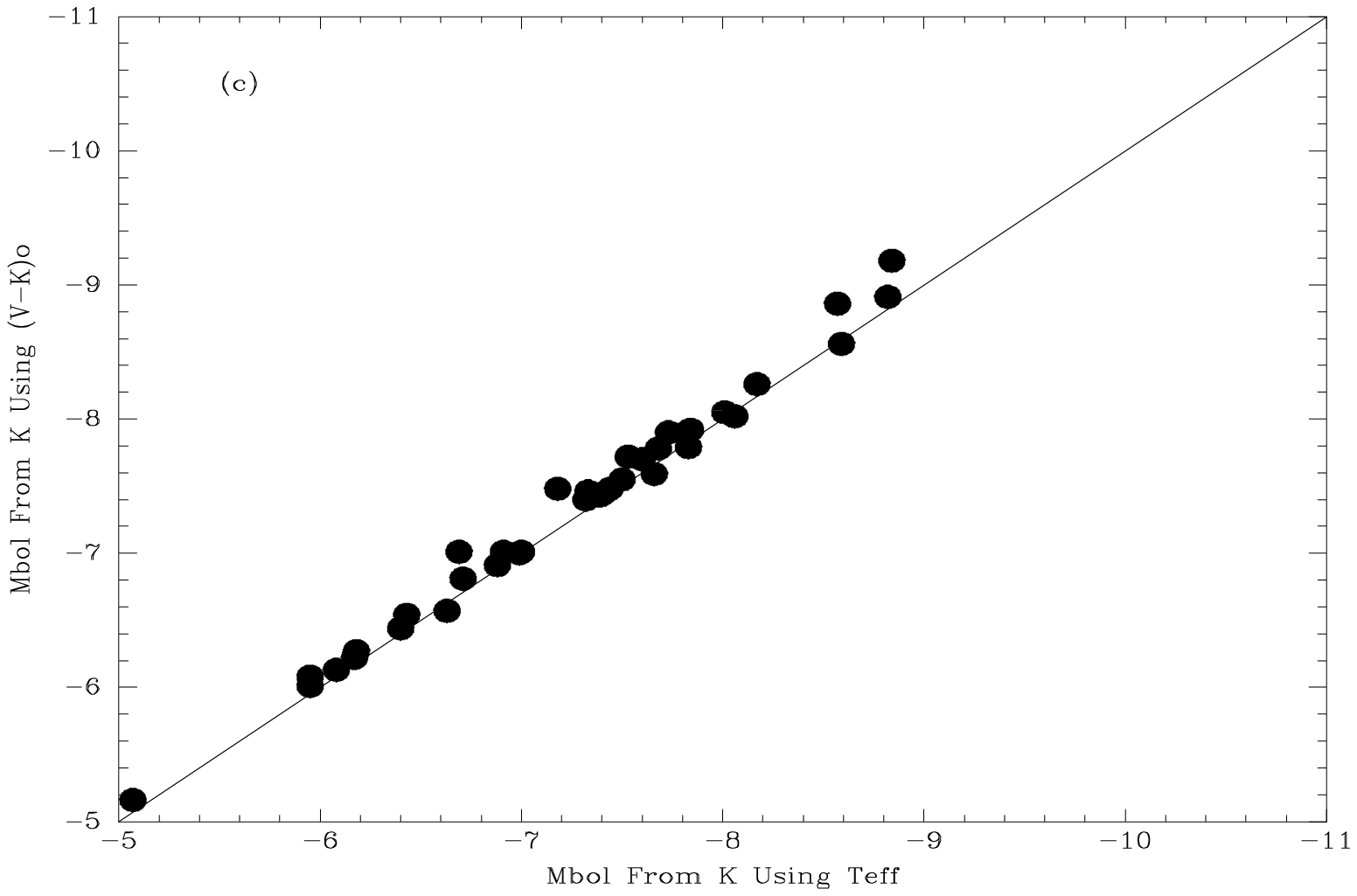}
\plotone{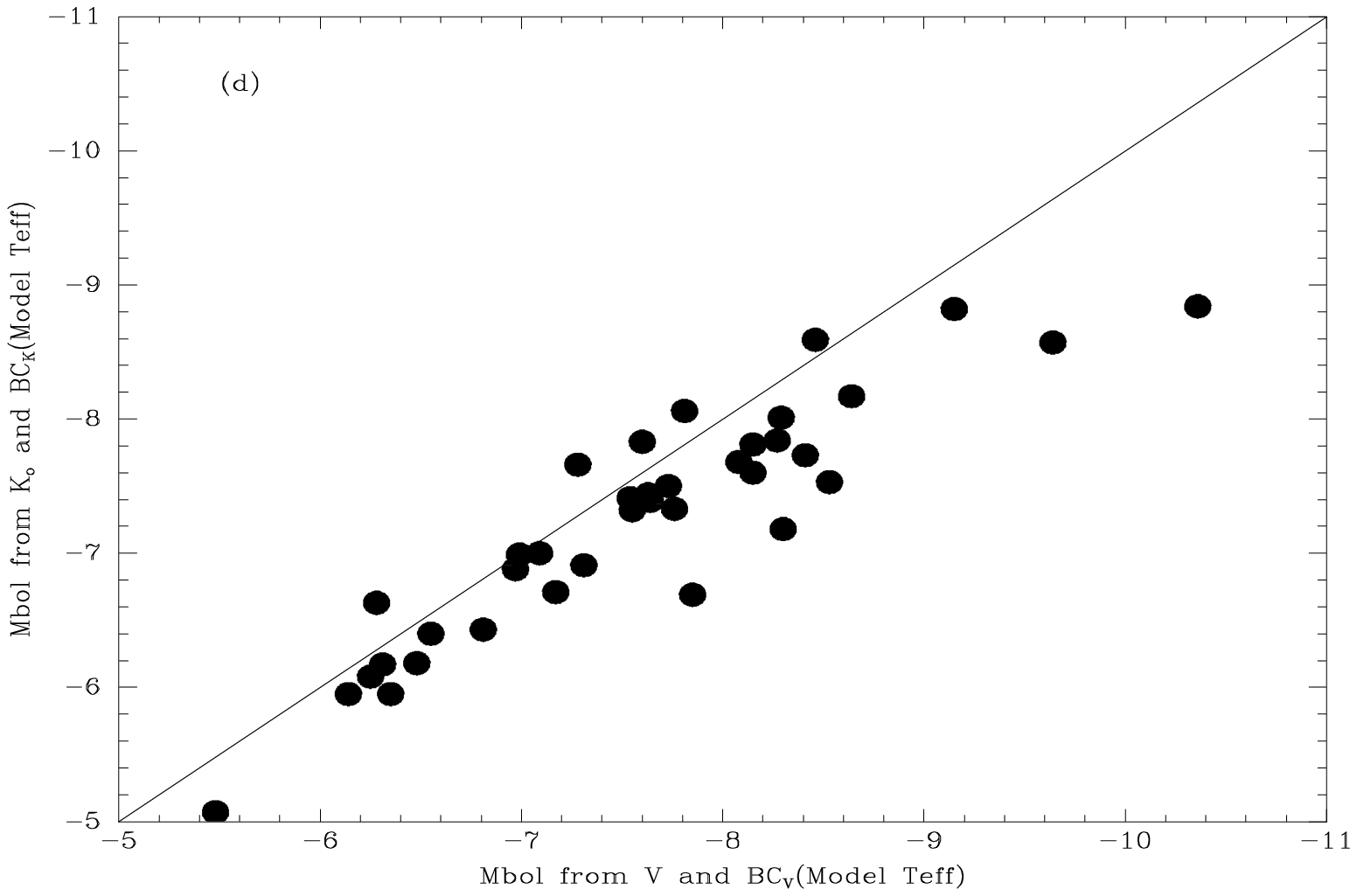}
\plotone{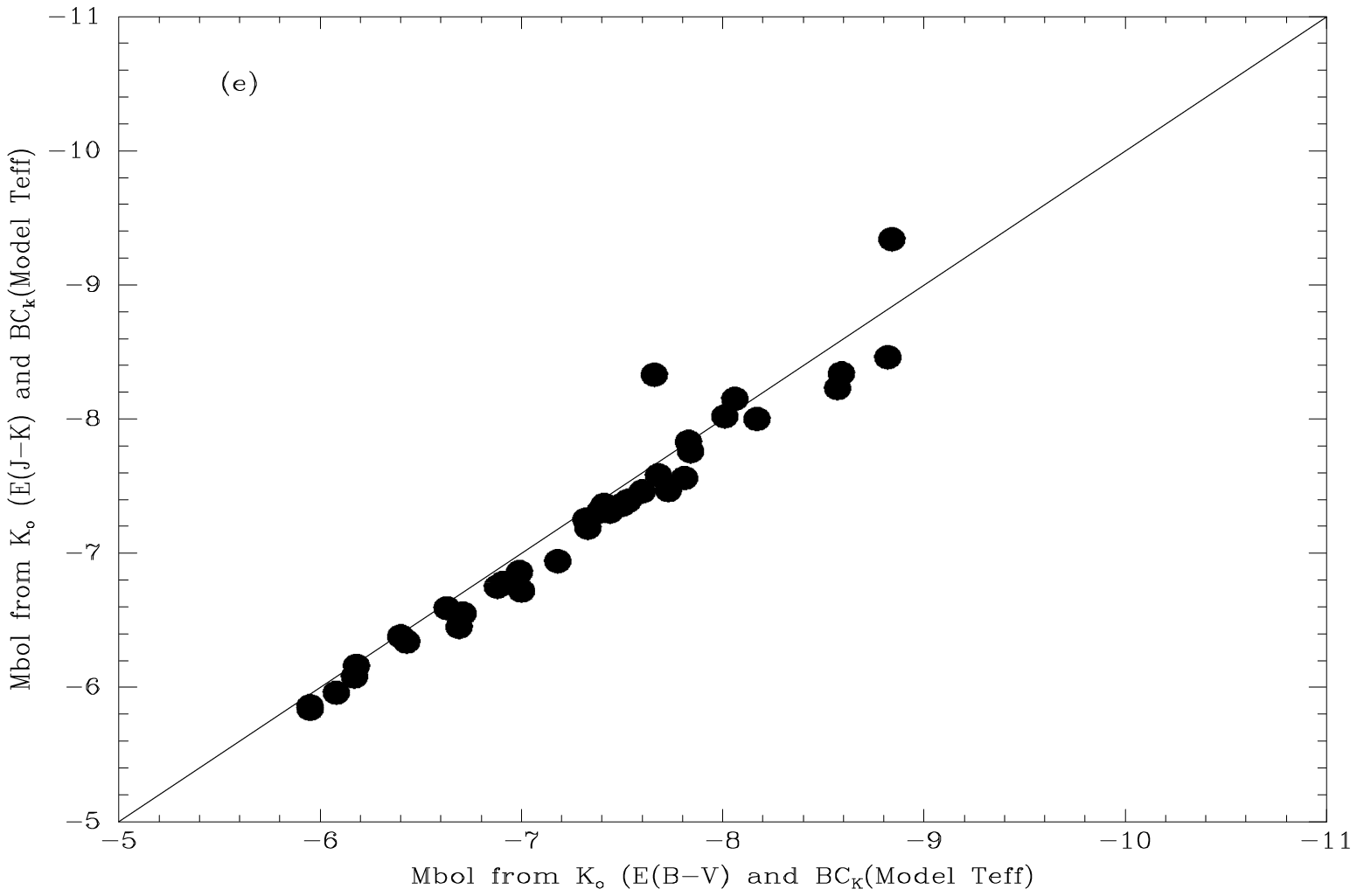}
\plotone{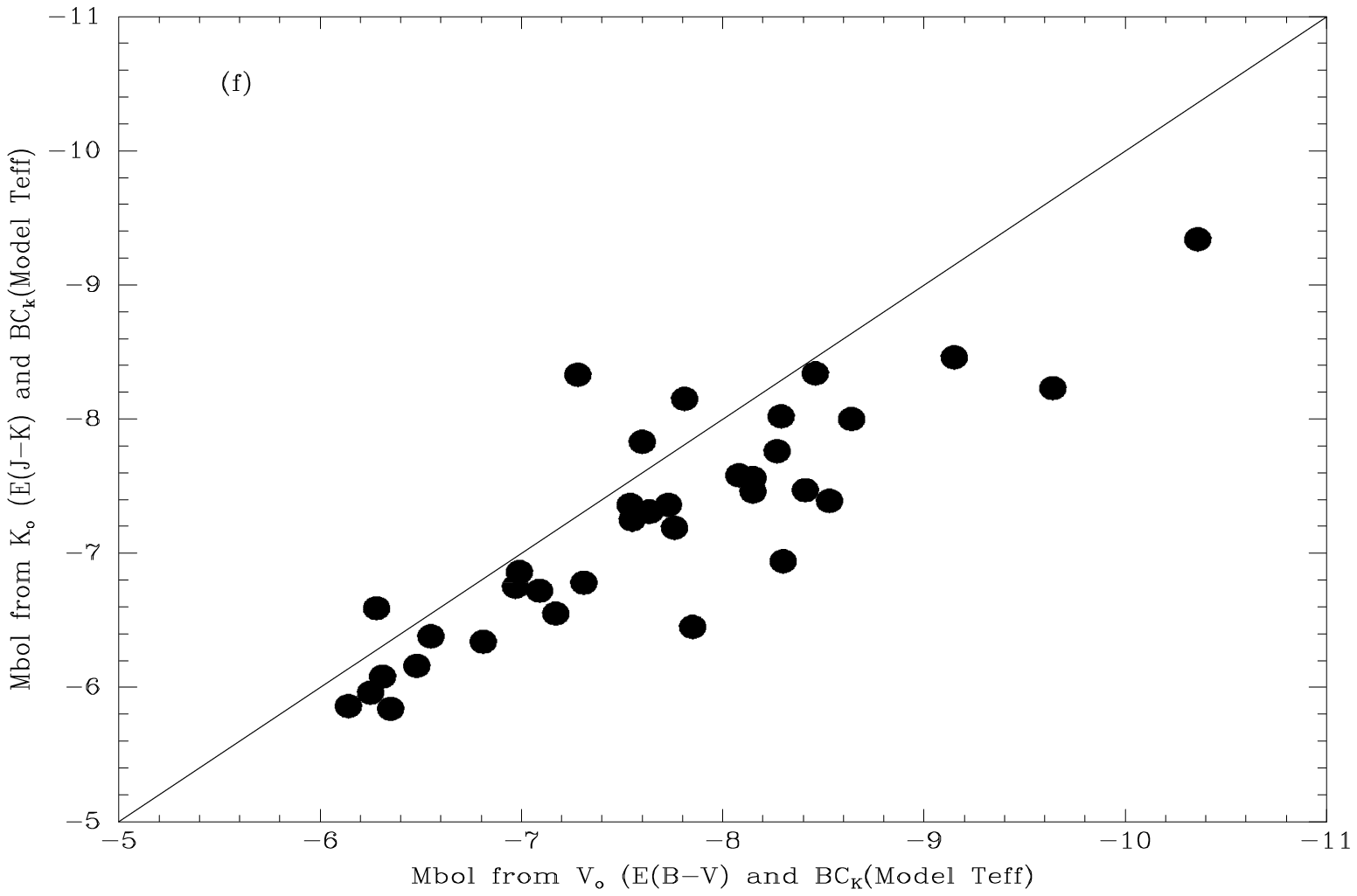}
\caption{\label{fig:K2} Derivation of the physical properties of RSGs
based upon K-band photometry  In (a) we compare the
effective temperatures derived from $(V-K)_o$ with those found by
fitting the models to our spectra.  The line shows the 1:1 relationship.
The $T_{\rm eff}$ values from the broad-band color shows considerable
scatter, and an average offset of 100~K.  
However, in (b) we find excellent 
agreement between the bolometric luminosities $M_{\rm bol}$ derived from
$V$ and our effective 
temperatures, and those derived from the $V-K$, in
accordance with the expectation of Josselin et al.\ (2000), who note the
usefulness of $K$ in deriving $M_{\rm bol}$.  In both (a) and (b) we
restrict the sample only to stars with $(V-K)_o$ colors between 2.9 and 8.0,
the ``sensible range" over which our transformation equations are good.
In (c) we demonstrate that deriving the bolometric luminosity from
the $(V-K)_o$ colors agrees well with deriving it from $K_o$ with the
BC at $K$ coming from the model fits in the optical.  In (d) 
we compare the bolometric luminosities derived from $K_o$ and the BC at $K$
coming from the model fits in the optical with the bolometric luminosities
derived purely from the optical fits.  Again, the agreement is excellent
for most stars, with a few outliers as noted in Table~\ref{tab:results}.
Finally, in (e) and (f)
we compare the bolometric luminosities at $K$ using $E(J-K)$ verses those
derived from $K$ (e) and $V$ (f) using our our $E(B-V)$'s.
The two outliers 
in (e) are KY Cyg
and KW Sgr, for which
the $J-K$ colors result in luminosities intermediate between our $V$ and
$K$ results.}
\end{figure}

We de-reddened the $V-K$ colors in Table~\ref{tab:stars} using the
extinction values derived from the model fits, and assuming
$E(V-K)=2.73 E(B-V)$, based on Schlegel et al.\ (1998); this does not
account for a modest change due to the
shifts of the effective wavelenths of the band-passes
for very red stars.
We compare the
effective temperatures derived by this method with those obtained from
the spectral types in Fig.~\ref{fig:K2}a.  The scatter is large, as one
might expect given the strong functional dependence of $T_{\rm eff}$ on
$(V-K)_o$, the variability of $V$, and the fact that the effective
temperatures now depend strongly upon the assumed reddenings. Note
that our uncertainty of 0.15 in A$_V$ translates to an
uncertainty of 0.13 in $E(V-K)$, and hence 50~K in $T_{\rm eff}$.  
However,
in general the agreement is good, with the $(V-K)_o$ colors yielding a
temperature whose median difference is 60 K 
warmer than our adopted scale. 

How well do the bolometric luminosities then agree?  In Fig.~\ref{fig:K2}b
we show the relationship between the bolometric luminosities derived from
$V$ and our $T_{\rm eff}$ values (Table~\ref{tab:results}) and those
found {\it purely from the de-reddened $V-K$ colors}.  The agreement here is
excellent, with only one significant outlier, KY~Cyg, the most luminous
star shown.

What if instead we had derived bolometric luminosities without reference
to the $V-K$ colors at all, but rather simply used the extinction
values from the fits to derive the absolute magnitude in the K-band 
[$M_K=K-0.37E(B-V)-(m-M)_o$], and then determined the bolometric correction
at K using the $T_{\rm eff}$ of the models fits?  We give that comparison
in Fig.~\ref{fig:K2}c.  Clearly there is excellent agreement.

In Fig.~\ref{fig:K2}d we show the comparison between the bolometric
magnitudes derived from the K-band in the same manner as in (c) and the
bolometric magnitudes derived from the visible.  This plot is similar
to that of Fig.~\ref{fig:K2}b, with excellent agreement in general.
There are five significant outliers,
whose 
differences are greater than 1 mag using the two methods 
(KY Cyg, CD-31 4916,
BD+35 4077, BD+60 2613, and HD 14528), all in the sense that the  bolometric
luminosity derived from $V$ may be overly luminous.  We flagged all five stars
in Fig.~\ref{fig:HRD} as well as in Table~\ref{tab:results}.

The fact that these five outliers were all 
more luminous based upon $V$ than upon $K$ raised the
the question as to whether our method was systematically
overestimating A$_V$. Recall that we did find that our
A$_V$ values tend to be higher than that of OB stars in the same
clusters and associations.  Following a suggestion offered by
the referee, we derived the bolometric luminosities at $K$ based instead
based upon the $J-K$ colors alone, 
ignoring our A$_V$ values.  We used the effective
temperatures derived from our model fits both to determine the bolometric
corrections, and to compute the intrinsic $(J-K)_o$ colors from a
relationship we found using the MARCS models:
$$(J-K)_o=3.10-0.547 (T_{\rm eff}/1000).$$  We adopt the broad-band
extinction terms suggested by Schlegel et al.\ (1998), i.e.,
$A_K=0.70 E(J-K)$.  Again, the numerical factor here does not take into
account the shift of $R_K$ due to the change in the
effective wavelengths of the broad-band filters, but it should be good
enough for the test we intend.
The $J-K$ photometry comes from Josselin et al.\ (2000)
and references therein. 
We show this comparison between
the two ways of deriving $M_{\rm bol}$ from $K$ in Fig.~\ref{fig:K2}e.  
There is excellent agreement between the two methods.  Thus, there cannot
be anything systematically wrong with our A$_V$ values in general. 
Interestingly, the two stars with the largest deviations in this figure are
KY Cyg and KW Sgr.  The $J-K$ correction suggests that if anything the
$M_{\rm bol}$ values should be intermediate between what we derive above
from $V$ and from $K$ using our values for the extinction, and that
perhaps our extinction values for these two stars are too low, rather
than too high. A comparison between the lumionsities derived from $K$
and $E(J-K)$ and those derived from $V$ and the extinctions derived
from our model fits is shown in Fig.~\ref{fig:K2}f; this is
very similar to what we found in Fig.~\ref{fig:K2}d.

We summarize the conclusions from these comparisons as follows.  First,
the MARCS models yield consistent results (to within 100~K) both from
the molecular band strengths and from the $V-K$ colors. Thus if one's
goal is simply to derive bolometric luminosities, $V$ and $K$ band data
will suffice for most purposes, if one has an estimate of the
reddenings. However, for accurate placement in the H-R diagram
(requiring $T_{\rm eff}$) then there is as yet no substitute for
spectroscopy and the use of molecular bands.  It is possible that other
colors, such as $V-R_c$, might prove more effective in determining the
effective temperatures than $V-K$, given the shorter baseline and
therefore lower sensitivity to reddening, i.e. $E(V-R_c)=0.64 E(B-V)$.
We will explore this further when we consider the Magellanic Cloud
sample, as these stars have simultaneous $V$ and $R$ measures, and the
reddenings are low and uniform (Massey et al.\ 1995).

\section{Conclusions and Summary}

We have determined a new effective temperature scale for Galactic RSGs by
fitting moderate resolution (4-6\AA) spectrophotometry with the new
MARCS stellar atmosphere models.  Our effective temperature scale is 
significantly warmer than previous scales, particularly for the later
M supergiants, where the differences amount to 400 K.  However,
our new results give  excellent agreement with the evolutionary tracks
of Meynet \& Maeder (2003), resolving the issue posed by Massey (2003)
that the evolutionary models do not produce RSGs as cool and as luminous
as ``observed".  

Our fitting showed excellent agreement between the models
and data for the majority of the stars.  The Ca~I $\lambda 4226$
and Ca~II H and K
atomic lines appear to be too strong in the models, but the molecular
transitions agree well at this dispersion. 
When we compare the physical properties derived from the model fits
to our optical spectrophotometry with those found from the models
using only the dereddened $V-K$ colors, we find good agreement,
providing an exacting demonstration of the self-consistency of the
MARCS models.

Extension of these studies to the Magellanic Clouds is underway, which should
help us understand the effect that metallicity has on the effective temperature
scale of RSGs.  In addition, the fact that the reddenings are small and uniform and the distances are known (van den Bergh 2000) will
allow further investigation
of colors as probes of the physical properties of these interesting massive 
stars.

\acknowledgements
This work was supported by NSF Grant AST 00-093060 
to PM, and
and the NSF's
Research Experience for Undergraduates program at Northern Arizona
University (AST 99-88007).
We are very grateful for the excellent hospitality and
support provided by the staff at KPNO and CTIO, and also thank 
Nat White and Kathy
Eastwood for their 
encouragement and guidance.  Hank Levesque pointed us towards 
some discussions of previous ``record holders" of the largest
stars known, and we also had
correspondence with Jim Kaler
on this topic. John Monnier kindly called his interesting results
on VY CMa to our attention.
We also gratefully acknowledge correspondence with Geoff Clayton.
Comments by an the referee, Roberta Humphreys, led to improvements in the 
presentation of our arguments in several places.

\begin{deluxetable}{l c c r r r r r l l l l}
\rotate
\tabletypesize{\scriptsize}
\tablewidth{0pc}
\tablenum{1}
\tablecolumns{12}
\tablecaption{\label{tab:stars}Program Stars}
\tablehead{
\colhead{Star}
&\colhead{$\alpha_{2000}$}
&\colhead{$\delta_{2000}$}
&\multicolumn{4}{c}{Photometry}
&\colhead{}
&\multicolumn{2}{c}{Spectral Type}
&\colhead{OB Association\tablenotemark{a}}
&\colhead{Comment} \\  \cline{4-7} \cline{9-10}
\multicolumn{3}{c}{}
&\colhead{$V$}
&\colhead{$B-V$}
&\colhead{$V-K$}
&\colhead{Ref\tablenotemark{b}}
&\colhead{}
&\colhead{Old\tablenotemark{c}}
&\colhead{New}
}
\startdata
BD+59 38        &00 21 24.29  &+59 57 11.2  & 9.13   &2.49  &  \nodata & 1&&M2 Iab      &M2 I    &Cas OB4         &MZ Cas           \\
Case 23         &00 49 10.71\tablenotemark{d} &+64 56 19.0\tablenotemark{d} &10.72   &2.77  & 7.80 & 2&&M1 Iab      &M3 I    &Cas OB7         &                 \\
HD 236697       &01 19 53.62  &+58 18 30.7  & 8.65   &2.16  & 5.41 & 3&&M2 Ib       &M1.5 I  &NGC 457         &V466 Cas         \\
BD+59 274       &01 33 29.19  &+60 38 48.2  & 8.55   &2.09  & 5.24 & 1&&M0 Ib       &M1 I    &Cas OB8/NGC581  &                 \\
BD+60 335       &01 46 05.48  &+60 59 36.7  & 9.15   &2.34  &  \nodata & 1&&M3 Iab      &M4 I    &Cas OB8/NGC663  &                 \\
HD 236871       &01 47 00.01  &+60 22 20.3  & 8.74   &2.27  &  \nodata & 2&&M3 Iab      &M2 I    &Cas OB8         &                 \\
HD 236915       &01 58 28.91  &+59 16 08.7  & 8.30   &2.20  &  \nodata & 2&&M2 Iab      &M2 I    &Per OB1-A       &                 \\
BD+59 372       &01 59 39.66  &+60 15 01.7  & 9.30   &2.28  &  \nodata & 2&&K5-M0 I     &K5-M0 I &Per OB1-A       &                 \\
BD+56 512       &02 18 53.29  &+57 25 16.7  & 9.23   &2.47  & 6.99 & 1&&M4 Ib       &M3 I    &Per OB1-D       &BU Per           \\
HD 14469\tablenotemark{e}       &02 22 06.89  &+56 36 14.9  & 7.63   &2.17  & 6.24 & 1&&M3-4 Iab    &M3-4 I  &Per OB1-D       &SU Per           \\
HD 14488        &02 22 24.30  &+57 06 34.3  & 8.35   &2.27  & 6.62 & 1&&M4 Iab      &M4 I    &Per OB1-D/NGC884&RS Per           \\
HD 14528        &02 22 51.72  &+58 35 11.4  & 9.23   &2.65  & 7.78 & 1&&M4e I       &M4.5 I  &Per OB1-D       &S Per            \\
BD+56 595       &02 23 11.03  &+57 11 58.3  & 8.18   &2.23  & 5.42 & 1&&M0 Iab      &M1 I    &Per OB1-D       &                 \\
HD 14580        &02 23 24.11  &+57 12 43.0  & 8.45   &2.27  & 5.43 & 2&&M0 Iab      &M1 I    &Per OB1-D       &                 \\
HD 14826        &02 25 21.86  &+57 26 14.1  & 8.24   &2.32  & 6.28 & 2&&M2 Iab      &M2 I    &Per OB1-D       &                 \\
HD 236979       &02 38 25.42  &+57 02 46.1  & 8.20   &2.35  & 6.17 & 4&&M2 Iab      &M2 I    &Per OB1-D?      &YZ Per           \\
W Per           &02 50 37.89  &+59 59 00.3  & 10.39  &2.53  & 8.30 & 4&&M3 Iab      &M4.5 I  &Per OB1-D?      &HD 237008        \\
BD+57 647       &02 51 03.95  &+57 51 19.9  & 9.52   &2.74  &  \nodata & 4&&M2 Iab      &M2 I    &Per OB1-D?      &HD 237010        \\
HD 17958        &02 56 24.65  &+64 19 56.8  & 6.24   &2.03  &  \nodata & 1&&K3 Ib       &K2 I    &Cam OB1         &                 \\
HD 23475\tablenotemark{e}       &03 49 31.28  &+65 31 33.5  & 4.48   &1.88  &  \nodata & 5&&M2+ IIab    &M2.5 II &\nodata             &                 \\
HD 33299        &05 10 34.98  &+30 47 51.1  & 6.72   &1.62  &  \nodata & 6&&K1 Ib       &K1 I    &Aur OB1         &                 \\
HD 35601        &05 27 10.22  &+29 55 15.8  & 7.35   &2.20  & 5.61 & 2&&M1 Ib       &M1.5 I  &Aur OB1         &                 \\
HD 36389\tablenotemark{e}       &05 32 12.75  &+18 35 39.2  & 4.38   &2.07  &  \nodata & 1&&M2 Iab-Ib   &M2 I    &\nodata             &                 \\
HD 37536\tablenotemark{f}       &05 40 42.05  &+31 55 14.2  & 6.21   &2.09  & 5.28 & 3&&M2 Iab      &M2 I    &Aur OB1         &                 \\
$\alpha$ Ori\tablenotemark{e}      &05 55 10.31  &+07 24 25.4  & 0.50   &1.85  &  \nodata & 1&&M1-2 Ia-Ib  &M2 I    &\nodata             &                 \\
HD 42475\tablenotemark{e}       &06 11 51.41  &+21 52 05.6  & 6.56   &2.25  & 5.70 & 3&&M0-1 Iab    &M1 I    &Gem OB1         & TV Gem          \\
HD 42543\tablenotemark{e}       &06 12 19.10  &+22 54 30.6  & 6.39   &2.24  & 5.46 & 3&&M1-2 Ia-Iab &M0 I    &Gem OB1         & BU Gem          \\
HD 44537\tablenotemark{e}       &06 24 53.90  &+49 17 16.4  & 4.91   &1.97  &  \nodata & 1&&K5-M0 Iab-Ib&M0 I    &\nodata             &                 \\
HD 50877\tablenotemark{e}       &06 54 07.95  &-24 11 03.2  & 3.86   &1.74  &  \nodata & 6&&K2.5 Iab    &K2.5 I  &Coll 121        &                 \\
HD 52005\tablenotemark{e}       &07 00 15.82  &+16 04 44.3  & 5.68   &1.63  &  \nodata & 3&&K3 Ib       &K5 I    &\nodata             &                 \\
HD 52877\tablenotemark{e}       &07 01 43.15  &-27 56 05.4  & 3.41   &1.69  & 3.90 & 6&&K7 Ib       &M1.5 I    &Coll 121        &$\sigma$ CMa        \\
CD-31 4916      &07 41 02.63  &-31 40 59.1  & 8.91   &2.16  & 5.20 & 1&&M2 Iab      &M2.5 I  &NGC2439         &                 \\
HD 63302\tablenotemark{e}       &07 47 38.53  &-15 59 26.5  & 6.35   &1.78  &  \nodata & 5&&K1 Ia-Iab   &K2 I    &\nodata             &                 \\
HD 90382        &10 24 25.36  &-60 11 29.0  & 7.45   &2.21  & 6.05 & 4&&M3 Iab      &M3-4 I  &Car OB1-D       &CK Car           \\
HD 91093        &10 29 35.37  &-57 57 59.0  & 8.31   &2.21  & 6.65 & 4&&M2 Iab      &M2 I    &Car OB1-A       &                 \\
CPD-57 3502\tablenotemark{e}    &10 35 43.71  &-58 14 42.3  & 7.44   &2.02  & 5.17 & 4&&M1.5 Iab-Ib &M1.5 I  &Car OB1-B/NGC329&                 \\
HD 303250       &10 44 20.04  &-58 03 53.5  & 8.92   &2.51  &  \nodata & 4&&M3 Iab      &M2 I    &Car OB1-B?      &                 \\
CD-58 3538\tablenotemark{e}     &10 44 47.15  &-59 24 48.1  & 8.36   &2.31  & 6.54 & 4&&M2+ Ia-0    &M2 I    &Car OB1-E       &RT Car           \\
HD 93420\tablenotemark{e}       &10 45 50.63  &-59 29 19.5  & 7.55   &1.87  & 6.15 & 4&&M4 Ib       &M4 I    &Car OB1-E       &BO Car           \\
HD 94096        &10 50 26.30  &-59 58 56.5  & 7.38   &2.24  & 5.64 & 4&&M2 Iab      &M2 I    &Car OB1-E       &IX Car           \\
HD 95687        &11 01 35.76  &-61 02 55.8  & 7.35   &2.12  & 5.81 & 4&&M2 Iab      &M3 I    &Car OB2         &                 \\
HD 95950        &11 03 06.15  &-60 54 38.6  & 6.75   &2.04  & 5.18 & 4&&M2 Ib       &M2 I    &Car OB2         &                 \\
HD 97671        &11 13 29.97  &-60 05 28.8  & 8.39   &2.52  & 7.42 & 4&&M3 Ia       &M3-4 I  &Car OB2\tablenotemark{g}     &                 \\
CD-60 3621      &11 35 44.96  &-61 34 41.0  & 7.27   &1.92  & 4.74 & 4&&M0 Ib       &M1.5 I  &NGC3766         &                 \\
HD 100930\tablenotemark{e}      &11 36 26.22  &-61 19 10.0  & 7.78   &1.95  & 5.68 & 4&&M2.5 Iab-Ib &M2.5 I  &\nodata             &                 \\
CD-60 3636      &11 36 34.84  &-61 36 35.1  & 7.62   &1.81  &  \nodata & 4&&M0 Ib       &M0 I    &NGC3766         &                 \\
V396 Cen        &13 17 25.05  &-61 35 02.3  & 7.85   &2.15  & 6.74 & 4&&M4 Ia-Iab   &M3-4 I  &Cen OB1-D       &HD 115283        \\
CPD-53 7344     &16 12 56.91  &-54 13 13.8  & 8.79   &1.78  &  \nodata & 4&&K2 Ib       &K2 I    &NGC6067         &                 \\
CPD-53 7364     &16 13 04.01\tablenotemark{d} &-54 12 21.2\tablenotemark{d} & 9.13   &1.86  &  \nodata & 4&&K4 Ib       &K2 I    &NGC6067         &                 \\
HD 160371\tablenotemark{e}      &17 40 58.55  &-32 12 52.1  & 6.14   &1.82  &  \nodata & 1&&K2.5 Ib     &K2.5 I  &M6              &BM Sco           \\
$\alpha$ Her\tablenotemark{e}      &17 14 38.86  &+14 23 25.2  & 3.06   &1.45  &  \nodata & 3&&M5 Ib-II    &M5 I    &\nodata             &                 \\
KW Sgr          &17 52 00.73  &-28 01 20.5  & 9.35   &2.78  & 7.98 & 4&&M3 Ia       &M1.5 I  &Sgr OB5         &HD 316496        \\
HD 175588\tablenotemark{e}      &18 54 30.28  &+36 53 55.0  & 4.30   &1.67  &  \nodata & 5&&M4 II       &M4 II   &\nodata             &$\delta^2$ Lyr        \\
HD 181475\tablenotemark{e}      &19 20 48.31  &-04 30 09.0  & 6.96   &2.14  &  \nodata & 6&&M0 II       &M1 II   &\nodata             &                 \\
HD 339034       &19 50 11.93  &+24 55 24.2  & 9.36   &3.05  &  \nodata & 1&&M1 Ia       &K3 I    &Vul OB1         &Case 15          \\
BD+35 4077      &20 21 12.37  &+35 37 09.8  & 9.72   &2.93  & 8.11 & 3&&M3 Iab      &M2.5 I  &Cyg OB1         &                 \\
BD+36 4025      &20 21 21.88  &+36 55 55.7  & 9.33   &2.49  & 8.75 & 3&&M3 Ia       &M3-4 I  &Cyg OB1         &BI Cyg           \\
BD+37 3903      &20 21 38.55  &+37 31 58.9  & 9.97   &3.26  & 9.75 & 3&&M3.5 Ia     &M3 I    &Cyg OB1         &BC Cyg           \\
KY Cyg          &20 25 58.08\tablenotemark{d} &+38 21 07.0\tablenotemark{d} &10.57   &3.64  &10.40 & 4&&M3 Ia       &M3-4 I  &Cyg OB1         &Case 66          \\
BD+39 4208\tablenotemark{e}     &20 28 50.59  &+39 58 54.4  & 8.69   &2.87  & 8.21 & 1&&M3-4 Ia-Iab &M3 I    &Cyg OB9         &RW Cyg           \\
HD 200905\tablenotemark{e}      &21 04 55.86  &+43 55 40.2  & 3.70   &1.65  &  \nodata & 5&&K4.5 Ib-II  &K4.5 I  &\nodata             &                 \\
HD 202380\tablenotemark{e}      &21 12 47.25  &+60 05 52.8  & 6.62   &2.39  &  \nodata & 2&&M2- Ib      &M2 I    &Cep OB2-A       &                 \\
HR 8248         &21 33 17.89  &+45 51 14.4  & 6.23   &1.78  &  \nodata & 7&&K4 Ib       &K1 I    &Cyg OB7        &HD 205349        \\
HD 206936\tablenotemark{e}      &21 43 30.46  &+58 46 48.1  & 4.08   &2.35  & 5.96 & 1&&M2 Ia       &M1 I    &Cep OB2-A       &$\mu$ Cep           \\
HD 210745\tablenotemark{e}      &22 10 51.28  &+58 12 04.5  & 3.35   &1.55  &  \nodata & 5&&K1.5 Ib     &K1.5 I  &\nodata             &                 \\
BD+56 2793      &22 30 10.73  &+57 00 03.1  & 8.09   &2.28  & 6.22 & 3&&M2 Ia       &M3 I    &Cep OB2-B       &HD 239978, ST Cep \\
Case 75         &22 33 35.0   &+58 53 45    &10.67   &3.18  &  \nodata & 4&&M1 Ia       &M2.5 I  &Cep OB1\tablenotemark{g}  &V354 Cep\tablenotemark{h}        \\
Case 78         &22 49 10.8   &+59 18 11    &10.76   &2.30  &  \nodata & 4&&M2 Ib       &M2 I    &Cep OB1\tablenotemark{g}  &V355 Cep\tablenotemark{h}        \\
HD 216946\tablenotemark{e}      &22 56 26.00  &+49 44 00.8  & 4.94   &1.77  &  \nodata & 5&&M0- Ib      &M0 I    &Lac OB1\tablenotemark{h}        &                 \\
Case 80         &23 10 10.90  &+61 14 29.9  & 9.72   &2.60  &  \nodata & 4&&M2 Iab      &M3 I    &Cas OB2         &GU Cep           \\
Case 81         &23 13 31.50\tablenotemark{d} &+60 30 18.5\tablenotemark{d} & 9.92   &2.70  &  \nodata & 4&&M2 Ia       &M2 I    &Cas OB2         &V356 Cep?        \\
HD 219978       &23 19 23.77  &+62 44 23.2  & 6.77   &2.27  &  \nodata & 2&&K5 Ib       &M1 I    &Cep OB3        &V809 Cas         \\
BD+60 2613      &23 44 03.28  &+61 47 22.2  & 8.50   &2.77  & 7.48 & 1&&M4 Ia       &M3 I    &Cas OB5         &PZ Cas           \\
BD+60 2634      &23 52 56.24  &+61 00 08.3  & 9.17   &2.51  & 7.22 & 3&&M2 Iab      &M3 I    &Cas OB5         &TZ Cas           \\
\enddata
\tablenotetext{a}{OB association membership from Humphreys 1978 and
Garmany \& Stencel 1992.
For Per OB1 and Car OB1, the sub-groups identified by
Mel'Nik \& Efremov 1995 have been used, based upon the star's $l$ and $b$.
}
\tablenotetext{b}{
References for $V$ and $B-V$: (1) Nicolet 1978; (2) Humphreys 1970; (3) Lee (1970); (4) Humphreys 1978 and references therein;
(5) Johnson et al.\ 1966; (6) Fernie 1983; (7) Jennens \& Helfer 1975. The
$K$ data come from Josselin et al.\ 2000 and references therein.}
\tablenotetext{c}{Old spectral types are from Humphreys 1978 and references
therein, except for the standard stars,  for which the types are from
Morgan \& Keenen.}
\tablenotetext{d}{Coordinates new to this study.}
\tablenotetext{e}{Spectral standard from Morgan \& Keenan 1973.}
\tablenotetext{f}{Possible AGB; see Josselin et al.\ 2003.}
\tablenotetext{g}{Membership listed as questionable by Humphreys 1978,
but see also Garmany \& Stencel 1992.}
\tablenotetext{h}{Incorrectly cross-referenced to the BD catalog by
Garmany \& Stencel 1992.}
\end{deluxetable}

\begin{deluxetable}{l c c r r r r r c c}
\tabletypesize{\scriptsize}
\tablewidth{0pc}
\tablenum{2}
\tablecolumns{10}
\tablecaption{\label{tab:dists}Adopted Distance Moduli and Average Reddenings}
\tablehead{
\colhead{OB Assoc./Cluster\tablenotemark{a}}
&\colhead{$l$}
&\colhead{$b$}
&\multicolumn{5}{c}{$(m-M)_o$ [mag]}
&\colhead{$E(B-V)_{\rm cluster}$}
&\colhead{Additional}  \\ \cline{4-8}
\multicolumn{3}{c}{}
&\colhead{Value}
&\colhead{Ref.\tablenotemark{b}}
&\colhead{Value}
&\colhead{Ref.\tablenotemark{b}}
&\colhead{Adopted}
&\colhead{[mag]}
&\colhead{$(m-M)_o$, (Ref.\tablenotemark{b})}
}
\startdata
Cas OB4           &119.5&-0.4&    11.0  &1   &12.3 &2   &11.6    &0.74 \\
Cas OB7           &123.5& 0.9&    12.0  &2   &\nodata  &\nodata   &12.0    &0.86 \\
NGC 457           &126.6&-4.4&    12.0  &2   &11.9 &3   &11.9    &0.47 \\
Cas OB8/NGC581    &128.0&-1.8&    11.9  &4   &11.7 &3   &11.8    &0.38 \\
Cas OB8/NGC663    &129.5&-1.0&    11.6  &4   &11.5 &3   &11.5    &0.78 \\
Cas OB8           &129.4&-0.9&    11.2  &1   &12.3 &2   &11.7    &0.70   \\
Per OB1-A         &131.1&-1.5&    11.0  &1   &11.8 &2   &11.4    &0.66      \\
Per OB1-D         &135.0&-3.5&    11.4  &1   &11.8 &2   &11.4    &0.66 \\
Per OB1-D/NGC 884  &135.1&-3.6&    12.0  &4   &11.9 &3   &11.9    &0.56 \\
Cam OB1           &140.4& 1.9&    10.0  &1   &10.0 &2   &10.0    &0.70 \\
Aur OB1           &174.6& 1.2&    10.7  &1   &10.6 &2   &10.7    &0.53 \\
Gem OB1           &188.9& 3.4&    10.6  &1   &10.9 &2   &10.7    &0.66 \\
Coll 121          &237.9&-7.7&     8.9  &5   & 8.4 &3   & 8.4    &0.03 \\
NGC 2439          &246.4&-4.4&    13.2  &2   &12.9 &3   &13.0    &0.41 \\
Car OB1-A         &284.5&-0.0&    11.9  &1   &12.0 &2   &11.9    &0.49 \\
Car OB1-B/NGC 3293 &285.9&+0.1&    12.0  &2   &11.8 &3   &11.9    &0.26 \\
Car OB1-B         &286.0& 0.5&    11.7  &1   &12.0 &2   &11.9    &0.26 \\
Car OB1-D         &286.6&-1.8&    11.9  &1   &12.0 &2   &11.7    &0.26 \\
Car OB1-E         &287.6&-0.7&    12.1  &1   &12.0 &2   &12.0    &0.26 \\
Car OB2           &290.6&-0.1&    11.7  &1   &11.5 &2   &11.6    &0.46 \\
NGC 3766          &294.1&-0.0&    11.6  &1   &11.7 &6   &11.6    &0.20 &11.2 (3) \\
Cen OB1-D         &305.5& 1.6&    11.4  &1   &12.0 &2   &11.6    &0.70 \\
NGC 6067          &329.8&-2.2&    11.6  &2   &10.8 &3   &10.8    &0.38 \\
M 6               &356.6&-0.7&     8.3: &2   & 8.4 &3   & 8.4    &0.14 \\
Sgr OB5           &  0.2&-1.3&    12.4  &2   & \nodata &\nodata   &12.4    &0.85  \\
Vul OB1           & 59.4&-0.1&    12.0  &1   &11.5 &2   &11.8    &0.83 &12.7 (7)\\
Cyg OB1           & 75.6& 1.1&    10.7  &1   &11.3 &2   &11.0    &0.97 \\
Cyg OB9           & 76.8& 1.4&    10.7  &1   &10.4 &2   &10.6    &1.08 \\
Cep OB2-A         & 99.3& 3.8&     9.9  &1   & 9.6 &2   & 9.7    &0.64 & 9.0 (5) \\
Cep OB2-B         &103.6& 5.6&     9.4  &1   & 9.6 &2   & 9.5    &0.64 & 9.0 (5)\\
Cep OB1           &108.5&-2.7&    11.7  &1   &12.7 &2   &12.2    &0.6:    \\
Cyg OB7           & 90.0& 2.0&     9.5  &2   & 9.6 &8   & 9.5    &0.4:\\
Lac OB1           & 96.8&16.1&     9.0  &1   & 8.9 &9   & 8.9    &0.11 & 7.8 (5) \\     
Cas OB2           &112.0& 0.0&    12.1  &2   & \nodata &\nodata   &12.1    &0.96 \\
Cep OB3           &110.4& 2.9&     9.6  &1   & 9.7 &2   & 9.7    &0.8: \\
Cas OB5           &115.5& 0.3&    11.8  &1   &12.0 &2   &11.9    &0.68 \\
\enddata
\tablenotetext{a}{Either classical names, or in the case of the sub-divided
associations, taken from Mel'Nik \& Efremov 1995.}
\tablenotetext{b}{References:
(1) Mel'Nik \& Efremov 1995;
(2) Humphreys 1978;
(3) WEBDA (Mermilliod \& Paunzen 2003);
(4) Becker \& Fenkart 1971;
(5) de Zeeuw et al 1999;
(6) Moitinho et al.\ 1997;
(7) Sagar \& Joshi 1981;
(8) Bochkarev \& Sitnik 1985;
(9) Ruprecht 1966.
}
\end{deluxetable}

\begin{deluxetable}{l c c c c c c}
\tabletypesize{\scriptsize}
\tablewidth{0pc}
\tablenum{3}
\tablecolumns{7}
\tablecaption{\label{tab:obs} Observation Parameters}
\tablehead{
\colhead{}
&\multicolumn{2}{c}{KPNO 2.1-m}
&\colhead{}
&\multicolumn{3}{c}{CTIO 1.5-m} \\ \cline{2-3} \cline{5-7}
\multicolumn{1}{l}{}
&\colhead{BLUE}
&\colhead{RED}
&\colhead{}
&\colhead{BLUE}
&\colhead{ORANGE}
&\colhead{RED}
}
\startdata
Grating/l mm$^{-1}$  &26/600  &58/400  & &26/600 &58/400 &58/400 \\
Blocking Filter &none &GG495 & &none &GG495 &OG570 \\
Wavelength Coverage(\AA\ ) &3200\tablenotemark{a} - 6000 &5000 - 9000 & &3500\tablenotemark{a} -5200 &5000 - 7500 &6300 - 9000 \\
Slit Width (''/$\mu$m) &3.0/250&2.1/170 & &4.9/270  &3.6/200 &3.6/200 \\
Dispersion(\AA\ mm$^{-1}$) &1.3 &1.9 & &1.5 & 2.2 &2.2 \\
Resolution(\AA) &3.6 &5.7 & &5.0 &6.4 &6.4 \\
\enddata
\tablenotetext{a}{Spectrophotometry below $\sim$4100\AA\ may be contaminated
by a grating problem for the reddest stars, and is not discussed further.}
\end{deluxetable}

\begin{deluxetable}{l l c c c c c c c c}
\tabletypesize{\scriptsize}
\tablewidth{0pc}
\tablenum{4}
\tablecolumns{10}
\tablecaption{\label{tab:results}Results of Model Fits}
\tablehead{
\colhead{Star}
&\colhead{Spectral}
&\colhead{$T_{\rm eff}$}
&\colhead{A$_V$}
&\multicolumn{2}{c}{$\log g$\tablenotemark{a}}
&\colhead{$R/R_o$\tablenotemark{a,b}}
&\colhead{$M_V$\tablenotemark{b}}
&\colhead{$M_{\rm bol}$\tablenotemark{a,b}}
&\colhead{$\Delta$A$_V$\tablenotemark{c}} \\ \cline{5-6}
\colhead{}
&\colhead{Type}
&\multicolumn{2}{c}{}
&\colhead{Model}
&\colhead{Actual\tablenotemark{d}}
}
\startdata
BD+59 38    &M2   I &3650&3.10& 0.0&   0.1& 600& -5.57& -7.17&0.81     \\                           
Case 23     &M3   I &3600&3.25& 0.5&   0.3& 410& -4.53& -6.28&0.59     \\                           
HD 236697   &M1.5 I &3700&1.55& 0.5&   0.4& 380& -4.80& -6.25&0.09     \\                           
BD+59 274   &M1   I &3750&1.55& 0.5&   0.4& 360& -4.80& -6.14&0.37    \\                            
BD+60 335   &M4   I &3525&2.63& 0.0&   0.1& 610& -4.99& -7.05&0.22     \\                           
HD 236871   &M2   I &3625&2.17& 0.0&   0.2& 520& -5.13& -6.80&0.00     \\                           
HD 236915   &M2   I &3650&1.71& 0.0&   0.3& 420& -4.80& -6.40&-0.34     \\                           
BD+59 372   &K5-M0 I&3825&2.48& 0.5&   0.6& 290& -4.58& -5.77&0.43    \\                            
BD+56 512   &M3 I   &3600&3.25& 0.5&   0.1& 620& -5.42& -7.17&1.21   \\                             
HD 14469    &M3-4 I &3575&2.01& 0.0&  -0.1& 780& -5.78& -7.64&-0.03     \\                           
HD 14488    &M4 I   &3550&2.63& 0.0&  -0.3&1000& -6.18& -8.15&0.90     \\                           
HD 14528    &M4.5 I &3500&4.18&-0.5&  -0.4/-0.1&1230/780& -6.36& -8.53/-7.53&2.14\\                 
BD+56 595   &M1 I   &3800&1.86& 0.0&   0.4& 380& -5.08& -6.31&-0.19    \\                            
HD 14580    &M1 I   &3800&2.17& 0.5&   0.4& 380& -5.12& -6.35&0.12    \\                            
HD 14826    &M2 I   &3625&2.48& 0.0&   0.0& 650& -5.64& -7.31&0.43     \\                           
HD 236979   &M2 I   &3700&2.32& 0.0&   0.2& 540& -5.52& -6.97&0.28     \\                           
W Per       &M4.5 I &3550&4.03& 0.0&   0.1& 620& -5.13& -7.09&1.98    \\                            
BD+57 647   &M2 I   &3650&4.03& 0.0&   0.0& 710& -5.91& -7.51&1.98     \\                           
HD 17958    &K2 I   &4200&2.17& 0.5&   0.5& 360& -5.93& -6.63&0.00    \\                            
HD 23475    &M2.5 II&3625&1.08& 0.0&   \nodata& \nodata&  \nodata &  \nodata &   \nodata\\          
HD 33299    &K1 I   &4300&0.77& 0.5&   0.9& 190& -4.76& -5.37&-0.87    \\                            
HD 35601    &M1.5 I &3700&2.01& 0.0&   0.2& 500& -5.36& -6.81&0.37     \\                           
HD 36389    &M2 I   &3650&1.24& 0.0&   \nodata& \nodata&  \nodata &  \nodata &   \nodata\\          
HD 37536    &M2 I   &3700&1.39& 0.0&   0.1& 630& -5.88& -7.33&-0.25     \\                           
$\alpha$ Ori&M2 I   &3650&0.62& 0.0&   \nodata& \nodata&  \nodata &  \nodata &   \nodata\\          
HD 42475    &M1 I   &3700&2.17& 0.0&  -0.1& 770& -6.31& -7.76&0.12  \\                              
HD 42543    &M0 I   &3800&2.01& 0.0&   0.0& 670& -6.32& -7.55&-0.03    \\                            
HD 44537    &M0 I   &3750&0.62& 0.0&   \nodata& \nodata&  \nodata &  \nodata &   \nodata\\          
HD 50877    &K2.5 I &3900&0.16& 0.5&   0.6& 280& -4.69& -5.75&0.06   \\                             
HD 52005    &K5 I   &3900&0.00& 0.0&   \nodata& \nodata&  \nodata &  \nodata &   \nodata\\          
HD 52877    &M1.5 I &3750&0.16& 0.5&   0.3& 420& -5.14& -6.48&0.06    \\                            
CD-31 4916  &M2.5 I &3600&2.01& 0.0&  -0.1/0.2& 850/500& -6.11& -7.85/-6.69&0.74   \\               
HD 63302    &K2 I   &4100&0.62& 0.0&   \nodata& \nodata&  \nodata &  \nodata &   \nodata\\          
HD 90382    &M3-4 I &3550&1.86& 0.0&  -0.3&1060& -6.31& -8.27&1.05     \\                           
HD 91093    &M2 I   &3625&2.01& 0.0&   0.0& 640& -5.60& -7.28&0.50    \\                            
CPD-57 3502 &M1.5 I &3700&1.08& 0.0&   0.2& 540& -5.54& -6.99&0.28    \\                            
HD 303250   &M2 I   &3625&2.94& 0.0&  -0.1& 750& -5.92& -7.60&2.14   \\                             
CD-58 3538  &M2 I   &3625&3.10& 0.0&  -0.3&1090& -6.74& -8.41&2.29    \\                            
HD 93420    &M4 I   &3525&1.08& 0.0&  -0.1& 790& -5.53& -7.60&0.28    \\                            
HD 94096    &M2 I   &3650&1.86& 0.0&  -0.2& 920& -6.48& -8.08&1.05     \\                           
HD 95687    &M3 I   &3625&1.71& 0.0&  -0.1& 760& -5.96& -7.63&0.28     \\                           
HD 95950    &M2 I   &3700&1.24& 0.0&   0.0& 700& -6.09& -7.54&-0.19    \\                            
HD 97671    &M3-4 I &3550&2.63& 0.0&  -0.2& 860& -5.85& -7.81&1.21     \\                           
CD-60 3621  &M1.5 I &3700&0.77& 0.0&   0.3& 440& -5.11& -6.55&0.16     \\                           
HD 100930   &M2.5 I &3600&1.08& 0.0&   \nodata& \nodata&  \nodata &  \nodata &   \nodata \\         
CD-60 3636  &M0 I   &3800&0.77& 0.5&   0.5& 320& -4.76& -5.98&0.16     \\                           
V396 Cen    &M3-4 I &3550&2.48& 0.0&  -0.3&1070& -6.33& -8.29&0.31     \\                           
CPD-53 7344 &K2 I   &4000&0.77& 1.0&   1.3& 100& -2.78& -3.70&-0.40     \\                           
CPD-53 7364 &K2 I   &4000&1.08& 1.0&   1.3& 100& -2.76& -3.67&-0.09     \\                           
HD 160371   &K2.5 I &3900&0.31& 1.0&   1.3& 100& -2.57& -3.63&-0.12     \\                           
$\alpha$ Her\tablenotemark{e}&M5 I &3450&1.40& 0.0&   \nodata& \nodata&  \nodata &  \nodata &   \nodata \\         
KW Sgr      &M1.5 I &3700&4.65&-0.5&  -0.5&1460& -7.70& -9.15&2.01    \\                            
HD 175588   &M4 II  &3550&0.47& 0.0&   \nodata& \nodata&  \nodata &  \nodata &   \nodata\\          
HD 181475   &M1 II  &3700&1.39& 0.0&   \nodata& \nodata&  \nodata &  \nodata &   \nodata\\          
HD 339034   &K3 I   &4000&5.27& 0.0&  -0.2& 980& -7.71& -8.63&2.70   \\                             
BD+35 4077  &M2.5 I &3600&5.27& 0.0&  -0.3/0.1&1040/620& -6.55& -8.30/-7.18&2.26   \\               
BD+36 4025  &M3-4 I &3575&5.11&-0.5&  -0.4&1240& -6.78& -8.64&2.11   \\                             
BD+37 3903  &M3 I   &3575&5.58&-0.5&  -0.3&1140& -6.61& -8.46&2.57    \\                            
KY Cyg      &M3-4 I &3500&7.75&-1.0&  -0.9/-0.5&2850/1420& -8.18&-10.36/-8.84&4.74  \\              
BD+39 4208  &M3 I   &3600&4.49& 0.0&  -0.2& 980& -6.41& -8.15&1.15     \\                           
HD 200905   &K4.5 I &3800&0.00\tablenotemark{f}& 0.0\tablenotemark{f}&\nodata& \nodata&  \nodata &  \nodata &\nodata \\               
HD 202380   &M2 I   &3700&2.63& 0.0&   0.1& 590& -5.72& -7.16&0.65     \\                           
HR 8248     &K1 I   &4000&0.93& 1.0&   0.9& 200& -4.20& -5.12&-0.31     \\                           
HD 206936   &M1 I   &3700&2.01&-0.5&  -0.5&1420& -7.63& -9.08&0.03  \\                              
HD 210745   &K1.5 I &4000&0.00& 0.0&   \nodata& \nodata&  \nodata &  \nodata &   \nodata \\         
BD+56 2793  &M3 I   &3600&2.32& 0.5&   0.6& 290& -3.73& -5.48&0.34    \\                            
Case 75     &M2.5 I &3650&6.05&-0.5&  -0.5&1520& -7.57& -9.17&4.18   \\                             
Case 78     &M2 I   &3650&4.65& 0.0&  -0.1& 770& -6.09& -7.69&2.79   \\                             
HD 216946   &M0 I   &3800&0.31& 0.5&   0.7& 260& -4.27& -5.50&-0.03  \\                              
Case 80     &M3 I   &3625&2.94& 0.0&   0.1& 570& -5.32& -7.00&-0.03  \\                              
Case 81     &M2 I   &3700&3.56& 0.0&   0.1& 590& -5.74& -7.19&0.59  \\                              
HD 219978   &M1 I   &3750&2.17& 0.5&   0.4& 410& -5.10& -6.44&-0.31    \\                            
BD+60 2613  &M3 I   &3600&4.49&-0.5&  -0.7/-0.4&1940/1190& -7.89& -9.64/8.57&2.39  \\               
BD+60 2634  &M3 I   &3600&3.25& 0.0&  -0.1& 800& -5.98& -7.73&1.15    \\                            
\enddata
\tablenotetext{a}{In the case of the five stars whose $M_{\rm bol}$ found
from $K$ differ significantly from that found from $V$, we list both values,
with the $V$ values first.}
\tablenotetext{b}{Given the 0.1~mag estimated error in $E(B-V)$, 
we expect that the $M_V$ and $M_{\rm bol}$ values are determined to 
approximately 0.3~mag.   Add to this the 50-100~K uncertainty in $T_{\rm eff}$,
the uncertainty in $R/R_\odot$ is roughly 20\%.}
\tablenotetext{c}{$\Delta$A$_V$=A$_V$ (RSG)-3.1 $E(B-V)_{\rm cluster}.$}
\tablenotetext{d}{Based upon adopting an approximate mass 
$\log (M/M_o) = 0.50 - .10 M_{\rm bol}$.
\tablenotetext{e}{The fit for $\alpha$ Her is poor in the blue, as shown in
the on-line figure, due presumably to the very close (0.2-arcsec) companion
known from interferometry. The star is listed as WDS 17146+1426 in
Mason et al.\ 2001.}
\tablenotetext{f}{The low reddening likely suggests a slightly higher
surface gravity would be appropriate for this standard.}
}
\end{deluxetable}

\begin{deluxetable}{l c l l c}
\tablewidth{0pc}
\tablenum{5}
\tablecolumns{3}
\tablecaption{\label{tab:NewT}New Effective Temperature Scale RSGs}
\tablehead{
\colhead{Spectral Type}
&\colhead{$T_{\rm eff}$\tablenotemark{a}}
&\colhead{$\sigma_{\mu}$}
&\colhead{N}
&\colhead{BC}
}
\startdata
K1-1.5 &4100 &100 &3 &-0.79 \\
K2-3   &4015 &40  &7 &-0.90 \\
K5-M0  &3840 &30  &3 &-1.16 \\
M0     &3790 &13  &4 &-1.25 \\
M1     &3745 &17  &7 &-1.35 \\
M1.5   &3710 &8   &6 &-1.43 \\
M2     &3660 &7   &17&-1.57 \\
M2.5   &3615 &10  &5 &-1.70 \\
M3     &3605 &4   &9 &-1.74 \\
M3.5   &3550 &11  &6 &-1.96 \\
M4-4.5 &3535 &8   &6 &-2.03 \\
M5     &3450 &\nodata  &1 &-2.49 \\
\enddata
\tablenotetext{a}{Averages from Table~\ref{tab:results}, rounded
to the nearest 5 K.}
\end{deluxetable}

\begin{deluxetable}{c c c}
\tablewidth{0pc}
\tablenum{6}
\tablecolumns{3}
\tablecaption{\label{tab:BCs}MARCS Bolometric Corrections\tablenotemark{a}}
\tablehead{
\colhead{$T_{\rm eff}$}
&\colhead{BC$_V$}
&\colhead{BC$_K$} 
}
\startdata
3200 &-4.58 & 3.16 \\
3300 &-3.66 & 3.08 \\
3400 &-2.81 & 3.00 \\
3500 &-2.18 & 2.92 \\
3600 &-1.75 & 2.84 \\
3700 &-1.45 & 2.76 \\
3800 &-1.23 & 2.68 \\
3900 &-1.06 & 2.61 \\
4000 &-0.92 & 2.53 \\
4100 &-0.79 & 2.47 \\
4200 &-0.70 & 2.40 \\
4300 &-0.61 & 2.33 \\
\enddata
\tablenotetext{a}{Computed for $\log g =0.0$ models, with the zero-point
defined such that
BC$_V=-0.07$ for the Sun; see Bessel et al.\ (1998).} 
\end{deluxetable}

\clearpage

\clearpage

\clearpage

\clearpage

\clearpage

\end{document}